\documentclass[aps,prb,onecolumn,nofootinbib,citeautoscript,10pt]{revtex4-2}

\usepackage{amsmath,amssymb,bm} 
\usepackage{graphicx,comment}

\usepackage[tight]{subfigure} 

\usepackage[dvipsnames]{xcolor} 
\usepackage[papersize={8.5in,11in}]{geometry}
\usepackage[colorlinks=true]{hyperref}
\hypersetup{
    bookmarks=true,         
    unicode=false,          
    pdftoolbar=true,        
    pdfmenubar=true,        
    pdffitwindow=false,     
    pdfstartview={FitH},    
    pdfkeywords={keyword1} {key2} {key3}, 
    pdfnewwindow=true,      
    colorlinks=true,       
    linkcolor=magenta, 
    citecolor=blue,        
    filecolor=magenta,      
    urlcolor=blue           
} 

\usepackage{graphicx}
\usepackage{dcolumn}
\usepackage{color}
\usepackage{amssymb,amsmath}
\usepackage{tabularx,graphicx}
\usepackage{epstopdf}
\usepackage{latexsym}
\usepackage{colortbl}
\usepackage{psfrag}
\usepackage{bbm,bm,array,physics}
\usepackage{dsfont}
\usepackage{float, mathrsfs}

\def \nn{\nonumber \\}
\def\*#1{\mathbf{#1}} 

\begin{document}
\title{Chiral anomaly and internode scatterings in multifold semimetals}

\author{Ipsita Mandal}
\email{ipsita.mandal@snu.edu.in}

\affiliation{Department of Physics, Shiv Nadar Institution of Eminence (SNIoE), Gautam Buddha Nagar, Uttar Pradesh 201314, India}

\begin{abstract} 
In our quest to unravel the topological properties of nodal points in three-dimensional semimetals, one hallmark property which  warrants our attention is the \textit{chiral anomaly}. In the Brillouin zone (BZ), the sign of the Berry-curvature field's monopole charge is referred to as the chirality ($\chi$) of the node, leading to the notion of \textit{chiral quasiparticles} sourcing \textit{chiral currents}, induced by internode scatterings proportional to the chiral anomaly. Here, we derive the generic form of the chiral conductivity when we have multifold nodes. Since the sum of all the monopole charges in the BZ is constrained to vanish, the nodes appear in pairs of $\chi =\pm 1$. Hence, the presence of band-crossing degeneracies of order higher than two make it possible to have two distinct scenarios: the pair of conjugate nodes in question comprise bands of (1) the same pseudospin variety and exhibiting Berry-curvature profiles differing by an overall factor of $\chi$, or (2) two distinct pseudospin representations. Covering these two possibilities, we apply our derived formula to semimetals harbouring triple-point (threefold-degenerate) and Rarita-Schwinger-Weyl (fourfold-degenerate) nodes, and show the resulting expressions for the conductivity featuring the chiral anomaly.
\end{abstract}

\maketitle

\tableofcontents

\section{Introduction}

There have been continuous efforts, both on the theoretical and experimental fronts, for unravelling the multifaceted exotic properties of three-dimensional (3d) semimetals, which harbour symmetry-protected band-crossing points in the Brillouin zone (BZ) near the Fermi level~\cite{burkov11_Weyl, yan17_topological, ips-kush-review, bernevig,ips-biref, grushin-multifold, ips-hermann-review, claudia-multifold}. Since the density-of-states goes exactly to zero at the nodal points, the semimetals differ from both the insulators (which always have a gap between the bands) and conventional metals (where bands overlap in finite regions of the BZ, with a finite density-of-states).
In general, the low-energy effective Hamiltonian of a system in the vicinity of a band-crossing point, with $(2\, \varsigma + 1) $ bands converging there, can be expressed as 
\begin{align}
\mathbf d( \mathbf{k} ) \cdot \boldsymbol{\mathcal{S}} , \quad 
\mathbf d( \mathbf{k} ) = \lbrace d_x ( \mathbf{k} ) , d_y ( \mathbf{k} ) , d_z ( \mathbf{k} )   \rbrace,
\end{align}
in the 3d momentum space ($\mathbf k  =\lbrace k_x, k_y, k_z  \rbrace $). Here, the vector operator $\boldsymbol{\mathcal{S}} \equiv \lbrace \mathcal S_x, \mathcal S_y, \mathcal S_z  \rbrace $ represents the three components of the angular momentum operator in the spin-$\varsigma$ representation of the SU(2) group. This is precisely the so-called $\mathbf k \cdot \mathbf p $ Hamiltonian, which can be obtained by performing \textit{ab initio} simulations for determining the bandstructures of the relevant materials. The emergent quasiparticles carry a set of quantum numbers, which we call pseudospin ($ \varsigma  $). The terminology of \textit{pseudospin} has been coined so as to unambiguously distinguish it (arising from crystal symmetries) from the relativistic spin (arising from the spacetime Lorentz invariance). 
In particular, the cases of multifold-band-crossing points have been identified in the 65 chiral space groups characterizing the chiral crystals~\cite{grushin-multifold}, which are the ones with only orientation-preserving symmetries. When the dominant terms involve linear-in-momentum dispersion, we have $ \mathbf d( \mathbf{k} )  = \mathbf k $. Such examples include the nodes of (1) a pseudospin-1/2 Weyl semimetal (WSM)~\cite{burkov11_Weyl, armitage_review, yan17_topological}, (2) a pseudospin-1 triple-point semimetal (TSM)~\cite{bernevig, ips3by2, ady-spin1, krish-spin1, ips-cd1, tang2017_multiple, grushin-multifold, prb108035428, ips-abs-spin1, claudia-multifold, ips-spin1-ph}, and (3) a pseudospin-3/2 Rarita-Schwinger-Weyl (RSW) semimetal~\cite{bernevig, long, igor, igor2, isobe-fu, tang2017_multiple, ips3by2, ips-cd1, ma2021_observation, ips-magnus, ips-jns, prb108035428, ips_jj_rsw, grushin-multifold, claudia-multifold, ips-rsw-ph, ips-shreya}. They involve two, three, and four bands crossing at the nodal point, respectively (as embodied in the value of $\varsigma $), and will constitute the example-systems that we will consider in this paper. It is interesting to note that, unlike the WSMs, integer-$\varsigma $ quasiparticles (with odd number of bands crossing at a nodal point) have no analogues in high-energy physics, since the occurrence of integer-spin relativistic fermions is naturally prohibited by the spin-statistics theorem. The wide range of transport signatures, that have been extensively investigated heretofore, include intrinsic anomalous-Hall effect~\cite{haldane04_berry,goswami13_axionic, burkov14_anomolous}, nonzero planar-Hall response \cite{zhang16_linear, chen16_thermoelectric, nandy_2017_chiral, nandy18_Berry, amit_magneto, das20_thermal, das22_nonlinear, pal22a_berry, pal22b_berry, fu22_thermoelectric, araki20_magnetic, mizuta14_contribution, ips-serena, ips_rahul_ph_strain, timm, rahul-jpcm, ips-kush-review, claudia-multifold, ips-ruiz, ips-tilted, ips-rsw-ph, ips-shreya}, magneto-optical conductivity under quantizing magnetic fields~\cite{gusynin06_magneto, staalhammar20_magneto, yadav23_magneto}, Magnus Hall effect~\cite{papaj_magnus, amit-magnus, ips-magnus}, circular dichroism \cite{ips-cd1, ips-cd}, circular photogalvanic effect \cite{moore18_optical, guo23_light,kozii, ips_cpge}, and transmission of quasiparticles across potential barriers/wells \cite{ips_aritra, ips-sandip, ips-sandip-sajid, ips-jns}. Such overwhelming literature has been fuelled by the fact that the 3d nodal-point semimetals are generically associated with nontrivial topology in the 3d momentum space, giving rise to vector fields in the forms of Berry curvature (BC) and orbital magnetic moment (OMM). These intrinsic topological quantities stem from the Berry phase \cite{xiao_review, sundaram99_wavepacket, timm, ips_rahul_ph_strain, graf-Nband, rahul-jpcm, ips-kush-review, claudia-multifold, ips-ruiz, ips-rsw-ph, ips-tilted, ips-shreya} and, when probed by externally-applied fields, lead to unique signatures by the virtue of affecting the response tensors.

For a nodal-point semimetal harbouring nontrivial topology, the nodes act as the singular points of the BC field, such that the converging/emanating bands carry BC monopoles \cite{fuchs-review, polash-review}, demarcating topological defects carrying nonzero topological charges. The sign of the monopole charge is referred to as the chirality $\chi$ of the node, leading to the notion of \textit{chiral} quasiparticles. We call them \textit{right-handed} or \textit{left-handed}, depending on whether $\chi = 1$ or $\chi = -1$. The sum of all the monopole charges, carried by either the conduction or the valence bands of all the chirally-charged nodes in the BZ, is constrained to vanish. This is in agreement with the Nielsen-Ninomiya theorem \cite{nielsen81_no}. Here, we adopt the convention that $\chi$ refers to the sign of the monopole charges of the negative-energy bands (i.e., the valence bands).\footnote{The nomenclature of ``conduction'' (positive-energy) and ``valence'' (negative-energy) bands refers to the signs of the dispersion, measured with respect to the nodal point (where we set the zero of energy).} Our convention implies that a positive (negative) sign indicates that the node acts as a source (sink) for the field lines of the BC.

In the context of high-energy physics, the relativistic Weyl fermions possess the hallmark property of the chiral anomaly, also known as the Adler-Bell-Jackiw anomaly of quantum electrodynamics \cite{adler, bell}. Strikingly, the phenomenon continues to hold in nonrelativistic settings involving WSMs \cite{chiral_ABJ, hosur-review, son13_chiral} and their higher-pseudospin generalizations (i.e., multifold nodal points) \cite{claudia-multifold}. In fact, for the low-energy physics governing condensed-matter systems, the anomaly reduces to the process of chiral-charge pumping from one node (with chirality $\chi $) to its conjugate (with chirality $ - \chi $), when we subject the material to external electric ($\mathbf E$) and magnetic ($\mathbf B$) fields. Since the rate of pumping is proportional to $  \mathbf E \cdot \mathbf  B $, we need $\mathbf{E} \cdot \mathbf{B} \neq 0$ to cause a net imbalance of chiral quasiparticles in the vicinity of an individual node. Of course, the total number of quasiparticles, obtained by summing over the conjugate pairs of nodes in the entire BZ, must yield zero, as required to conserve the net electric charge. In other words, a net chiral current must appear as a purely quantum-mechanical effect (due to $\mathbf{E} \cdot \mathbf{B} \neq 0$), although the total particle current must vanish.

\begin{figure*}[t]
{\includegraphics[width=0.6\linewidth]{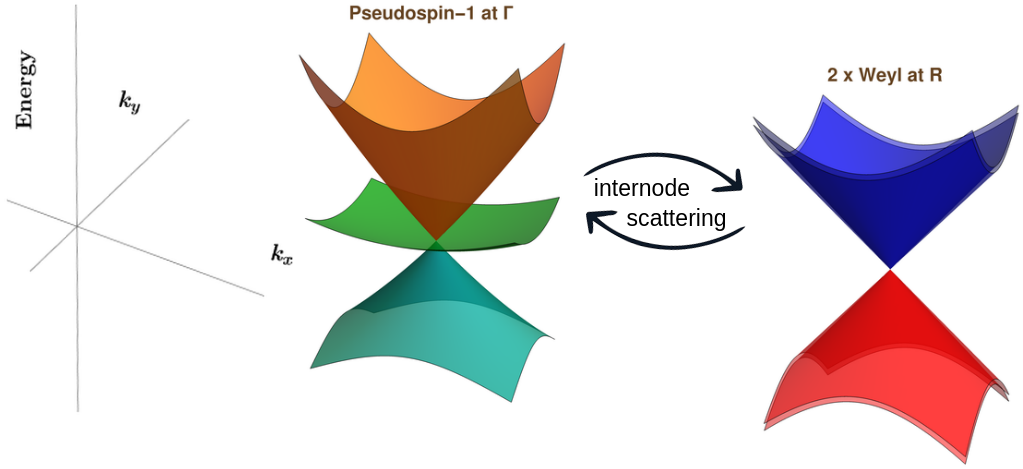}}
\caption{\label{figdis1}Schematics of the multiple bands of a single pseudospin-1 triple-point node (at the $ \Gamma $-point) and a double-pseudospin-1/2 node (at the $ R $-point) against the $k_x k_y$-plane. The internode scatterings, thus, involve bands having distinct topological properties, although the monopole charge (summed over either the conduction or the valence bands) from one node is the negative of the other.
}
\end{figure*}

\begin{figure*}[t]
{\includegraphics[width=0.6\linewidth]{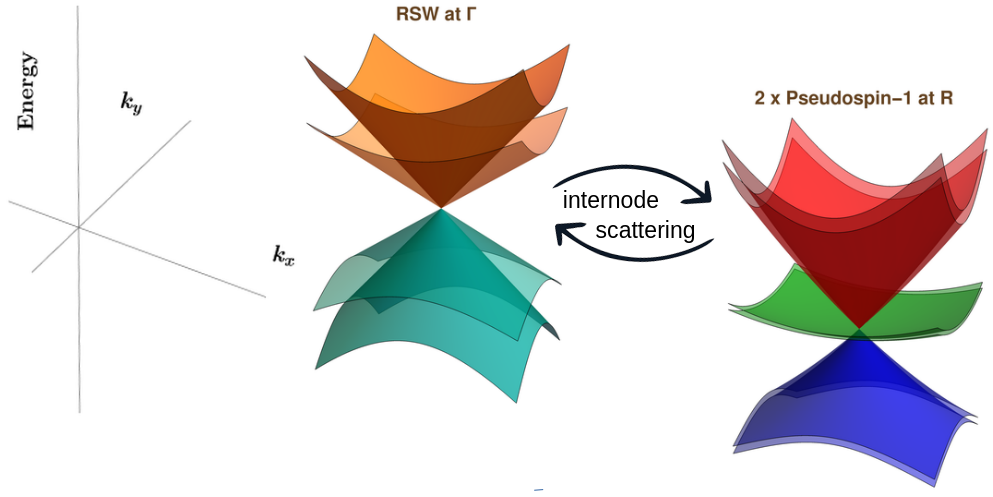}}
\caption{\label{figdis2}Schematics of the multiple bands of a single RSW node (at the $\Gamma$-point) and a double-pseudospin-1 triple-point node (at the $ R $-point) against the $k_x k_y$-plane. The internode scatterings, thus, involve bands having distinct topological properties, although the monopole charge (summed over either the conduction or the valence bands) from one node is the negative of the other.
}
\end{figure*}

In this paper, we apply the semiclassical Boltzmann formalism \cite{ips-kush-review, ips_rahul_ph_strain, ips-rsw-ph}, using the relaxation-time approximation, to determine the chiral current induced by the chiral anomaly, in the regime of nonquantizing magnetic fields. This involves considering a collision integral ($ I_{\text{coll}} $) which contains a part ($I_{\text{coll}}^{\text{inter}}$) induced by internode scatterings. While earlier works \cite{yip, amit_magneto} have considered this problem (specifically, for WSMs), the issue of scattering between nodes of multifold nodal points has not been addressed. For multifold degeneracies, we encounter two distinct situations: (1) the pair of conjugate nodes in question are of the same pseudospin variety and the bands at the two nodes bands have BC profiles differing by an overall factor of $\chi$; (2) the pair of conjugate nodes comprise bands of different pseudospin quantum numbers (and, thus, automatically having distinct BC fields). The first case is exemplified by a pair of TSMs \cite{bernevig, ady-spin1}. The second case is exemplified by the following two realizations:
\\(a) A single node of TSM is pinned at the center of the BZ (i.e., the $\Gamma $-point), carrying a monopole charge of $+ \, 2 $, while a fourfold-degenerate node (comprising two copies of WSMs of the same chirality) exists at the boundary of the BZ (i.e., the $R$-point) with a net monopole charge equalling $- \,1- 1 = - \, 2$. This is shown schematically in Fig.~\ref{figdis1}. Candidate materials include CoSi \cite{claudia-multifold}.
\\(b) In a typical material harbouring an RSW node \cite{tang2017_multiple, prb108035428, prl119206401, yamakage}, we find that there is the RSW node at the $\Gamma $-point carrying $+ \, 4 $ charge, and a sixfold-degenerate (originating from the doubling of pseudospin-1 excitations) at the $ R $-point carrying $- \, 4 $ charge. This is shown schematically in Fig.~\ref{figdis2}. Candidate materials include the SrGePt family (e.g., SrSiPd, BaSiPd, CaSiPt, SrSiPt, BaSiPt, and BaGePt) \cite{prb108035428}.
\\The bottomline is that, in realistic materials, multifold band-crossings, in general, may not arise in conjugate pairs \cite{claudia-multifold, grushin-multifold, tang2017_multiple} belonging to the same pseudospin representation --- the Nielson-Ninomiya theorem \cite{nielsen81_no} is satisfied by the net monopole charge being zero on considering all the nodal points in the entire BZ. Hence, computing the internode-scattering-contribution from such nodes of different nature involves severe conceptual challenges, which we address here by deriving a generic formula. In this context, we would like to point out that, in Ref.~\cite{claudia-multifold}, the authors have not resorted to a rigorous derivation for the internode-scattering part, and have taken an expression with phenomenological parameters.

The paper is organized as follows: In Sec.~\ref{secboltz}, we describe the derivation of the generic expression of the conductivity tensor ($  {\sigma}^{\chi, \rm inter }_{s} $), which is connected to the chiral-charge pumping induced by the chiral anomaly. Sec.~\ref{secexample} is devoted to finding the explicit expressions for $ {\sigma}^{\chi, \rm inter }_{s} $, considering some specific systems (described above). Finally, we end with a summary and some future perspectives in Sec.~\ref{secsum}. We provide the details of various expressions/intermediate steps in the appendices. Throughout the paper, we will be using natural units.

\section{Boltzmann formalism}
\label{secboltz}

Let us consider the transport by quasiparticles, in the vicinity of the node with chirality $\chi $, for a 3d nodal-point semimetal. Let us define the distribution function by $ f_s^\chi ( \mathbf r , \mathbf k, t) $ for the quasiparticles occupying a Bloch band labelled by the index $s$, with the crystal momentum $\mathbf k$, band-dispersion $ \varepsilon^\chi_s (\mathbf k)$, and group velocity $ \boldsymbol{v}^{\chi}_{s}  (\mathbf k) =  
\nabla_{\mathbf{k}} \varepsilon^\chi_s  (\mathbf k) $. In general, where we do not have nodes of the same nature for $\chi$ and $ -\chi $, $s$ is a function of $\chi $ [i.e., $s = s (\chi)$]. However, to unclutter the notations, we suppress the $\chi$-dependence of the $s$-index, and it will be clear from the $\chi$-index of the relevant quantities what values of $s$ we are talking about.
If $ \mathbf \Omega^{\chi}_{s}  (\mathbf k) $ is the associated BC, then
\begin{align}
\mathcal{D}^{\chi}_{s}  (\mathbf k) =  \frac{1}
{ 1 + \, e  \left [  \mathbf{B} \cdot
\mathbf \Omega^{\chi}_{s}  (\mathbf k) \right ] }
\end{align} 
is the factor which modifies the phase-space volume element from $ dV_p (\mathbf r , \mathbf k) \equiv \frac{ d^3 \mathbf k}{(2\, \pi)^3 } 
\,d^3 \mathbf r $ to $  \left[ \mathcal{D}^{\chi}_{s}   (\mathbf k) \right ]^{-1} \, dV_p (\mathbf r , \mathbf k) $, such that the Liouville’s theorem (in the absence of collisions) continues to hold in the presence of a nonzero BC \cite{son13_chiral, xiao05_berry, duval06_Berry, son12_berry}. Hence, the modified classical probability-density function, centered at $\left \lbrace \mathbf r , \mathbf k \right \rbrace $ at time $t$, is given by
\begin{align}
\label{eqdn}
dN_s^\chi (\mathbf r , \mathbf k) =  g_s^\chi  
\left[ \mathcal{D}^{\chi}_{s} (\mathbf k) \right ]^{-1} \,
f_s^\chi ( \mathbf r , \mathbf k, t) \, dV_p (\mathbf r , \mathbf k)\,.
\end{align}
Here, $ g_s^\chi $ denotes the degeneracy of the band.

Let us define
\begin{align}
\xi^\chi_s (\mathbf k) = \varepsilon^\chi_s  (\mathbf k) 
+ \eta^\chi_s  (\mathbf k)\,, \quad
\eta^\chi_s  (\mathbf k) = -\, \mathbf{m}^{\chi}_{s}( \mathbf k) 
\cdot {\mathbf B} \,, \quad
\boldsymbol{w}^{\chi}_{s} = \boldsymbol{v}^{\chi}_{s} 
+ \boldsymbol{u}^\chi_s  (\mathbf k) \,, \quad
\boldsymbol{u}^\chi_s  (\mathbf k)
= \nabla_{\mathbf{k}} \eta^\chi_s  (\mathbf k)\,,
\end{align}
where $\mathbf{m}^{\chi}_{s}( \mathbf k) $ is the OMM and $ \eta^\chi_s  (\mathbf k) $ is the OMM-induced correction to the effective dispersion. Similarly, $\boldsymbol{ w }^\chi_s  (\mathbf k) $ is the OMM-corrected the band velocity. Further details of the application of the Boltzmann formalism have been reviewed and explained in Appendix~\ref{appboltz}.

Using the explicit forms of the solutions for $f_s^\chi$~\cite{nandy_2017_chiral, amit_magneto}, the contribution to the electric-current density, from the band with index $s$, is captured by
\begin{align}
\label{eqcur}
{\mathbf J}^\chi_s
& =   -\, e \,   g_s^\chi   \int
\frac{ d^3 \mathbf k}{(2\, \pi)^3 } \,
(\mathcal{D}^{\chi}_{s})^{-1}    \, \dot{\mathbf r}
\,  f_s^\chi( \mathbf r , \mathbf k) 
\Rightarrow {\mathbf J}^\chi_s
 =   -\, e \,   g_s^\chi   \int
\frac{ d^3 \mathbf k}{(2\, \pi)^3 } \,
\left[   \boldsymbol{w}^{\chi}_{s} + e \, ({\mathbf E}  \cross  
\mathbf \Omega^{\chi}_{s})  + e   \, ( \mathbf \Omega^{\chi}_{s} \cdot 
\boldsymbol{w}^{\chi}_{s} ) \, \mathbf B  \right]
\,  f_s^\chi( \mathbf r , \mathbf k) \, .
\end{align}
Henceforth, we will set $ g_s^\chi  =1 $, assuming nondegenerate bands (away from the nodal points where the bands cross) and ignoring any electronic spin-degeneracy.

\subsection{Conductivity arising from internode scatterings}
\label{secintersigma}

We include the internode scatterings in a relaxation-time approximation, where we treat the internode-scattering time $\tau_G$ as a phenomenological constant (analogous to the intranode-scattering time $\tau$). To start with, let us assume that initially, in the infinite past (denoted by time $ t = -\infty $), both the nodes had the same chemical potential $E_F$, characterized by the distribution function
\begin{align}
f_0 (\varepsilon  ) =\frac{1} {1 + e^{\frac{\varepsilon - E_F} {T}}} \,,
\end{align} 
in the absence of any externally applied fields.
Eventually, on applying the electromagnetic fields, there is the onset of the chiral anomaly, causing the two nodes to acquire a local equilibrium value of chemical potential, given by $\mu_\chi$. The concepts of local and global equilibria have been introduced in Refs.~\cite{son13_chiral, amit_magneto, deng2019_quantum}.
More details can be found in Appendix~\ref{appinter}.

We define the average over all the possible electron states (which reduces to the average over all the possible electron states at the Fermi level at $T=0$) of a physical observable
$\mathcal O^\chi_s ( \xi^\chi_s (\mathbf k) , \mu, T) $ as
\begin{align}
\bar{\mathcal{O}}_\chi \equiv
\left \langle \mathcal{O}^\chi_s ( \xi^\chi_s (\mathbf k), E_F ,T) 
\right  \rangle = 
\frac{
\sum \limits_s \int \frac{d^3 \mathbf{k}} {(2 \pi)^3} 
\left(  {\mathcal D}^\chi_{s} (\mathbf k) \right)^{-1}
\left [ - f_0^\prime (\xi^\chi_s (\mathbf k) ) \right ] 
\mathcal{O}^\chi_s ( \xi^\chi_s (\mathbf k), E_F ,T) }
{ \sum \limits_{\tilde s} \int 
\frac{d^3 \mathbf{ q} }  {(2 \pi)^3} 
\left( {\mathcal D}^\chi_{\tilde s}  (\mathbf q) \right)^{-1}
\left [ - f_0^\prime (\xi^\chi_{\tilde s} (\mathbf{ q}) ) \right ]  } \,,
\end{align}
borrowing the notation introduced in Ref.~\cite{deng2019_quantum}.
Let the symbol
\begin{align}
\label{eqdos}
\rho_\chi \equiv   \sum \limits_{s} \int 
\frac{d^3 \mathbf{k} }  {(2 \pi)^3} \left( {\mathcal D}^\chi_{s} (\mathbf k) \right)^{-1}
\left [ - f_0^\prime (\xi^\chi_{s} (\mathbf{k}) ) \right ]
\end{align}
represent the density-of-states at node $\chi $ [cf. Eq.~\eqref{eqdn}].
Finally, the conductivity tensor, corresponding to the internode-scattering-induced current, is given by
\begin{align}
\label{eqcond1}
& \left( {\sigma}^{\chi, \rm inter }_{s} \right)_{ij} =
\frac{  e^2  \, \rho_{-\chi} }  
{ \rho_\chi \, \rho_G  }  
\left [  \tau_G
-\frac{\tau \, \rho_G } { \rho_{-\chi}}   \right ]
 \int
\frac{ d^3 \mathbf k}{(2\, \pi)^3 } 
\left [- f_{0}^\prime (\xi^{\chi}_{s}) \right ] 
\;  \left[   \left ( {w}^{\chi}_{s} \right)_i
+ e   \, ( \mathbf \Omega^{\chi}_{s} \cdot 
\boldsymbol{w}^{\chi}_{s} ) \,  B_i  \right]\,
\, {\mathcal I}_j^\chi  \,, \quad
\rho_G = \frac{ \rho_\chi + \rho_{-\chi} }  {2} \,,
\nn &
\boldsymbol {\mathcal I}^\chi  = \rho_\chi
\left \langle
\mathcal{D}^{\chi}_{s} 
\left \lbrace {\boldsymbol{w}}_s^\chi 
+ e \left({\mathbf \Omega}^\chi_{s} \cdot {\boldsymbol{w}}_s^\chi   \right)  
\mathbf B \right \rbrace\right  \rangle 
=
\sum \limits_s \int \frac{d^3 \mathbf{k}} {(2 \pi)^3} 
\left [ - f_0^\prime (\xi^\chi_s (\mathbf k) ) \right ] 
 \left \lbrace {\boldsymbol{w}}_s^\chi 
+ e \left({\mathbf \Omega}^\chi_{s} \cdot {\boldsymbol{w}}_s^\chi   \right)  
\mathbf B \right \rbrace.
\end{align}

We show the expanded form of the integrand in Appendix~\ref{appexp}, retaining terms upto order $ B^2 $.
For the cases when $ \varepsilon^\chi_s $ is a function of magnitudes of the momentum-components [i.e., $ \varepsilon^\chi_s (\mathbf k) =  \varepsilon^\chi_s (|k_x|, |k_y|, |k_z|)$], $ f_0^{\prime}\big (  \varepsilon_s^\chi (\mathbf k) \big) $ and its derivatives (with respect to $ \varepsilon^\chi_s $) are even functions of $\mathbf k $. Since integrands which are odd functions of the momentum-components must vanish, we conclude that
\begin{align}
\label{eqiso}
& \left( {\sigma}^{\chi, \rm inter }_{s} \right)_{ij} = 
\frac {e^2  
\left [ \tau_G \, \rho_{-\chi}^{(0)} 
- \tau \, \rho_G^{(0)} \right ]
\Upsilon^{\chi, s}_i \, {\mathcal  I}_j^{\chi, 1} }
 { \rho_G^{(0)} \, \rho_{\chi}^{(0)} } 
 + \order{B^3} \,,
\quad \rho_{\chi}^{(0)} =
 \sum \limits_{ \tilde s }
\int \frac{ d^3 \mathbf q}  {(2\, \pi)^3 } 
 \left \lbrace  - f_0^\prime 
 \big (\varepsilon^\chi_{ \tilde s } 
 (\mathbf q) \big)  \right \rbrace ,\quad
\rho_G^{(0)} = 
\frac{ \rho_\chi^{(0)} + \rho_{-\chi}^{(0)} }  {2}   \,, \nn 
& \mathcal I^{\chi, 1}_j = 
 \sum_{\tilde s} \Upsilon^{\chi, \tilde s}_j \,,
\quad \Upsilon^{\chi, s}_j
=   B_j \int
\frac{ d^3 \mathbf k} {(2\, \pi)^3 } 
\left [
 e \, {\mathbf \Omega}_s^{\chi} (\mathbf k) 
 \cdot  {\boldsymbol  v}_s^{\chi} (\mathbf k) 
\left \lbrace - 
 f_0^\prime \big (\varepsilon_s^{\chi} (\mathbf k) \big)  
 \right  \rbrace
+
  \left ( m_s^{\chi} (\mathbf k) \right)_j  
  \left (  v_s^{\chi} (\mathbf k) \right)_j
f_0^{\prime \prime} \big (\varepsilon_s^{\chi} (\mathbf k)  \big) 
\right ] .
\end{align}
We note that the first nonzero term is of the order $B^2$, which agrees with the results found in the literature \cite{son13_chiral, amit_magneto, yip} for the case of a pair of conjugate nodes for a WSM (harbouring the simplest case of twofold nodes). Needless to say that our formula of course applies to the generic cases of multifold nodes.

For the special case when we have scatterings between two nodes of the same nature, with no energy offset between their nodal points, Eq.~\eqref{eqiso} further simplifies to
\begin{align}
\label{eqsamenode}
 \left( {\sigma}^{\chi, \rm inter }_{s} \right)_{ij}  
 & =
 \frac { e^2   \left (\tau_G - \tau \right) }
{ \rho_1^{(0)} } \,
 \Upsilon^{1, s}_i \, \sum \limits_{\tilde s} \Upsilon^{1, \tilde s}_j\,,
\end{align}
where we have used the facts that $s (\chi ) = s (-\chi ) $, $\varepsilon^\chi_s = \varepsilon^{-\chi}_s \equiv \varepsilon_s $, $   \rho_\chi^{(0)}=  \rho_{-\chi}^{(0)} $, $ v^\chi_s = v^{-\chi}_s $, $ {\mathbf \Omega}_s^{\chi} =  
-\, {\mathbf \Omega}_s^{ -\chi} $, and $ {\boldsymbol m}_s^{\chi} =  
-\, {\boldsymbol m}_s^{ -\chi} $.

\subsection{Comparison with earlier works}

For internode scatterings between two nodes involving pseudospin-1/2 representation and the same magnitude of the Chern number (e.g., two conjugate nodes of the WSM variety), we might compare the treatment outlined in Refs.~\cite{yip, amit_magneto} with our generic formula presented above. They both agree qualitatively, which we explain here.
Firstly, Refs.~\cite{yip, amit_magneto} have used a characteristic internode-scattering time, $ \tau_X $, defined  from $\tau_{12}$ and $\tau_{21}$, which denote the internode relaxation times considering node 1 (i.e., $\chi = +1 $) and node 2 (i.e., $\chi = -1$), respectively. The relations are given by
$$ \frac{1} {\tau_{12} } = \frac{1} {\tau_X} \,
 \frac{\sqrt{D_1 \, D_2 } }  { {\tilde D}_1 (B) }
\text{ and }
\frac{1} {\tau_{21} } = \frac{1}{\tau_X} \, \frac{\sqrt{D_1 \, D_2 } }
{\tilde D_2 (B) }.
$$
Here, $\tilde D_1 (B) = \rho_1 $ and $\tilde D_2 (B) = \rho_{-1} $ in terms of the notation $\rho_\chi $, that we have used to indicate the of density-of-states in this paper [cf. Eq.~\eqref{eqdos}]. Furthermore, $ D_1 (B) \equiv \tilde D_1 (B = 0)$ and $ D_2 (B) \equiv \tilde D_2 (B = 0)$.
Thus, $ \tau_X $ represents some kind of geometric mean (involving the density-of-states) which can be used to characterize the relaxation time for either node. Secondly, Refs.~\cite{yip, amit_magneto} have resorted to describing the global equilibration process by arguing that the internode scatterings will tend to force the local value of the chemical potential (1) $\mu_1 $ towards $\mu_2 \equiv \mu_{-1}$ for node 1; and (2) $\mu_2 $ to $\mu_1 $ for node 2. Comparing with our treatment, we have argued instead that the internode scatterings will tend to bring the value of $\mu_\chi $ to $\mu_G$, with the latter representing the chemical potential averaged over the two nodes. This we find a more reasonable argument to use.
Consequently, we have used the parameter $\tau_G $, which is characteristic of either node, irrespective of its individual band-crossing nature. In the end, this leads to minute quantitative differences in the final formulae. In fact, both approaches lead to the same qualitative conclusions for twofold nodal points, if we remember that both $\tau_X $ and $\tau_G$ are phenomenological parameters in the limit of the relaxation-time approximation.

\section{Specific examples}
\label{secexample}

Using the generic forms derived in Sec.~\ref{secintersigma}, we now aim to derive the explicit expressions of the conductivity tensors embodying the chiral currents in specific systems. For the sake of simplicity, we will limit ourselves to the $ T \rightarrow 0 $ limit [such that $ \left \lbrace  -  f_0^{\prime}  \big (  \varepsilon_s^\chi (\mathbf k) \big) \right \rbrace \rightarrow \delta \big ( \varepsilon_s^\chi (\mathbf k) - E_F \big) $]. We consider the situations when both $\mu_\chi >0$ and $\mu_{-\chi} > 0$, such that the chiral quaiparticles occupying the positive-energy bands participate in transport. For all the Hamiltonians shown below, $v_0$ denotes the effective Fermi velocity for the isotropic and linearly dispersing bands. In order to carry out the integrations, taking advantage of the isotropy of the systems that we consider here, we resort to using the spherical polar coordinates (as shown in Appendix~\ref{secint}). We assume $E_F > 0 $ in what follows:

\subsection{Weyl semimetal}

A single node of a WSM, with chirality $\chi $, is represented by
\begin{align}
	\mathcal{H}^\chi_{\rm WSM}(\mathbf{k}) =  v_0\left( 
k_x\,	{\sigma }_x + k_y\, {\sigma }_y
+ \chi \, k_z \, {\sigma }_z \right) - \Theta(-\chi)\, \Delta,
\end{align}
where $ \boldsymbol{\sigma } = \lbrace {\sigma}_x,\, {\sigma }_y,\, {\sigma}_z \rbrace $ represents the vector operator comprising the three Pauli matrices. Here, we have set the zero of the energy at the nodal point of the node with $\chi=+1$, and have kept a generic energy offset $-\Delta  $ for the conjugate node with $\chi=-1$. 
The dispersion relations of the two bands meeting at a nodal point are given by
\begin{align}
\label{eqeval1}
\varepsilon_s^{\rm{WSM},\chi} ( k ) =  s \, v_0 \, k 
- \Theta(-\chi)\, \Delta\,, \quad 
s =\pm 1,
\end{align}
where $ k = \sqrt{k_x^2 + k_y^2 + k_z^2 } $.
The corresponding group velocities of the quasiparticles are given by 
\begin{align} 
\boldsymbol{v}_s^{\rm WSM}(\mathbf{k}) &= 
\nabla_{\mathbf{k}}  \varepsilon_s^{\rm WSM} (\mathbf{k})  
=  \frac{ s \, v_0 \,\mathbf k }{k} \, .  
\end{align}
The BC and the OMM are captured by the following expressions:
\begin{align}  
{\mathbf \Omega}^{\chi}_{s}( \mathbf k) &=    
	 - \, \frac{\chi \,   s \,\mathbf k  } { 2\, k^3}  \text{ and } 
\mathbf{m}^{\chi}_{s}( \mathbf k) 
=  -\frac{e \, \chi\, v_0 \, \mathbf k } { 2\, k^2}  \,.
\end{align}

For $\Delta = 0$, we can use Eq.~\eqref{eqsamenode} to determine the internode-scattering-induced conductivity components, considering a pair of conjugate Weyl nodes. It reduces to the simple form of
\begin{align}
\label{eqsigweyl}
 \left( {\sigma}^{\chi, \rm inter }_{s=1} \right)_{ij}  
 & = \frac {   2\, e^4\, v_0^3 \left (\tau_G - \tau \right)
   B_i \, B_j}
{ 9 \, \pi^2\, E_F^2} \,.
\end{align}
Summing over the nodes with $\chi =\pm 1$, the net conductivity tensor is simply captured by $  4 \, e^4\, v_0^3 \left (\tau_G - \tau \right)
   B_i \, B_j /( 9 \, \pi^2\, E_F^2)  $.

Choosing $\Delta>0 $, only the conduction band at each node contribute. Using Eq.~\eqref{eqiso}, we obtain
\begin{align}
\left( {\sigma}^{1, \rm inter }_{s=1} \right)_{ij}  
 & = \frac {2 \, e^4 \, v_0^3 \, B_i \, B_j}
{9\, \pi^2 \, E_F^2}
 \left[
  \frac {2\left ( E_F + \Delta \right)^2\tau  _G}
    {\Delta^2 + 2 \, E_F\left ( E_F + \Delta \right)}
   - \tau  \right ] , \nn
\left( {\sigma}^{-1, \rm inter }_{s=1} \right)_{ij}  
 &= 
  \frac {2 \, e^4 \, v_0^3 \, B_i \, B_j}
{9\, \pi^2 \, (E_F + \Delta)^2 }
 \left[
\frac {2 \, E_F^2 \, \tau  _G}
    {\Delta^2 + 2 \, E_F \left ( E_F + \Delta \right)}
   - \tau  \right ].
\end{align}
$\sum_\chi \left( {\sigma}^{\chi, \rm inter }_{s=1} \right)_{ij} $ gives the net value.

\subsection{Triple-point semimetal}

A single node of a TSM, with chirality $\chi $, is represented by
\begin{align}
\label{eqhamspin1}
\mathcal{H}^\chi_{\rm TSM}(\mathbf  k) = v_0 \left(k_x \,\mathcal S_x + k_y \,\mathcal S_y 
+ \chi \,  k_z \,\mathcal S_z \right) 
-  \Theta(-\chi)\, \Delta,
\end{align}
where $ \boldsymbol{\mathcal S} =  \lbrace {\mathcal S }_x,\, {\mathcal S }_y,\, {\mathcal S }_z \rbrace$ represents the angular-momentum vector operator in the spin-$ 1$ representation. We choose
\begin{align}
\mathcal{S}_x = \frac {1} {\sqrt{2}}
\begin{pmatrix}
0&1&0\\1&0&1\\0&1&0
\end{pmatrix} ,\quad
\mathcal{S}_y =\frac{1}{\sqrt{2}}
\begin{pmatrix}
0&- i &0\\ \mathrm{i} &0&-  i \\
0&  i  &0
\end{pmatrix},\quad
\mathcal{S}_z =
\begin{pmatrix}
1&0&0\\0&0&0\\0&0&-1
\end{pmatrix} .
\end{align}
The energy eigenvalues, group velocities, BC, and OMM are obtained in the following forms:
\begin{align}
\label{eqeval3}
& \varepsilon_s^{\rm{TSM},\chi} ( k ) =  s \, v_0 \, k 
- \Theta(-\chi)\, \Delta
\text{ and }
\boldsymbol{v}_s^{\rm TSM} (\mathbf{k}) = 
\nabla_{\mathbf{k}}  \varepsilon_{s}(\mathbf{k})  
=  \frac{ s \, v_0 \,\mathbf k }{k} \,,
\text{ where } 
s \in  \left \lbrace  \pm 1, 0 \right \rbrace; \nn
& {\mathbf \Omega}^{\chi}_{s}( \mathbf k) =    
	 - \, \frac{\chi \,   s \,\mathbf k  } { k^3}  \text{ and } 
\mathbf{m}^{\chi}_{s}( \mathbf k) 
= - \frac{e \, \chi\, v_0 \, \mathcal{G}_s  \,\mathbf k } { 2\, k^2} \, ,
\text{ where } \lbrace \mathcal{G}_{\pm 1}, \,\mathcal{G}_0 \rbrace
 =  \lbrace 1 \,,  2  \rbrace \, .
\end{align}

For the internode-scattering-induced conductivity components with $\Delta = 0$, arising from a pair of conjugate nodes of a TSM \cite{bernevig, ady-spin1}, Eq.~\eqref{eqsamenode} gives us
\begin{align}
 \left( {\sigma}^{\chi, \rm inter }_{s=1} \right)_{ij}  
 & = \frac { 49 \, e^4\, v_0^3 \left (\tau_G - \tau \right)
   B_i \, B_j}
{ 72 \, \pi^2\, E_F^2} \,.
\end{align} 
Summing over the nodes with $\chi =\pm 1$, the net conductivity tensor is simply captured by $  49 \, e^4\, v_0^3 \left (\tau_G - \tau \right)
   B_i \, B_j /( 36 \, \pi^2\, E_F^2)  $.
We would like to point out that the overall numerical factor differs from that of the WSM case [cf. Eq.~\eqref{eqsigweyl}], which is expected, because the BC vector differs by a factor of two.

For $\Delta >0 $, Eq.~\eqref{eqiso} gives us
\begin{align}
\left( {\sigma}^{1, \rm inter }_{s=1} \right)_{ij}  
 & = \frac { 49 \, e^4 \, v_0^3 \, B_i \, B_j}
{ 72\, \pi^2 \, E_F^2}
 \left[
  \frac {2\left ( E_F + \Delta \right)^2\tau  _G}
    {\Delta^2 + 2 \, E_F\left ( E_F + \Delta \right)}
   - \tau  \right ] , \nn
\left( {\sigma}^{-1, \rm inter }_{s=1} \right)_{ij}  
 &= 
  \frac { 49 \, e^4 \, v_0^3 \, B_i \, B_j}
{ 72\, \pi^2 \, (E_F + \Delta)^2 }
 \left[
\frac { 2 \, E_F^2 \, \tau  _G}
    {\Delta^2 + 2 \, E_F \left ( E_F + \Delta \right)}
   - \tau  \right ].
\end{align}
$\sum_\chi \left( {\sigma}^{\chi, \rm inter }_{s=1} \right)_{ij} $ gives the net value.

For the case depicted in Fig.~\ref{figdis1}, let us consider the internode-induced conductivity for the chiral quasiparticles at the TSM node with $\chi =1 $ (occupying the band $s=1$). Let us assume that the $\chi=-1$ node has an energy offset of $- \Delta $ (where $\Delta > 0$) with respect to the $\chi= 1$ node. Applying Eq.~\eqref{eqiso} then leads to the node-specific conductivity-tensor components of
\begin{align}
 \left( {\sigma}^{1, \rm inter }_{s=1} \right)_{ij}  
 & = \frac {49 \, e^4\, v_0^3\, B_i\, B_j}
{72 \, \pi^2\, E_F^2}
\left [\frac {4\, v_0^3\left ( E_F + \Delta \right)^2
       \tau  _G}
    { E_F^2\left (\tilde {v}_0^3 + 2 \, v_0^3 \right)
+ 2\,\Delta \, v_0^3 \left( 2\, E_F + \Delta \right)
}
   - \tau  \right]
  \text{ at the $\Gamma$-point and}\nn
 2\times \left( {\sigma}^{-1, \rm inter }_{s=1} \right)_{ij}  
 & = 
\frac { 4 \, e^4\, \tilde{v}_0^3\, B_i\, B_j}
{ 9 \, \pi^2\left(E_F + \Delta \right )^2 }
\left [ \frac {2 \, E_F^2 \, \tilde {v}_0^3 \, \tau_G}
    { E_F^2\left (\tilde {v}_0^3 + 2 \, v_0^3 \right)
+ 2\,\Delta \, v_0^3 \left( 2\, E_F + \Delta \right)
}
   - \tau  \right]
  \text{ at the $R$-point}.
\end{align} 
Here, $v_0$ and $\tilde v_0 $ denote the group velocities of the pseudospin-1 (for TSM) and pseudospin-1/2 (for WSM) quasiparticles, respectively. Summing over the two nodes, we get the total value as
$
 \left( {\sigma}^{1, \rm inter }_{s=1} \right)_{ij}   +
 2\times \left( {\sigma}^{-1, \rm inter }_{s=1} \right)_{ij}  \,.$

\subsection{Rarita-Schwinger-Weyl semimetal}

The explicit form of the Hamiltonian for a single RSW node with chirality $\chi $ is given by 
\begin{align}
	\mathcal{H}_{\rm RSW}(\mathbf{k}) =  v_0\left( 
k_x\,	{\mathcal J }_x + k_y\, {\mathcal J }_y
+ \chi \, k_z \, {\mathcal J }_z \right),
\end{align}
where $ \boldsymbol{\mathcal J } = \lbrace {\mathcal J }_x,\, {\mathcal J }_y,\, {\mathcal J }_z \rbrace $ represents the vector operator whose three components comprise the the angular momentum operators in the spin-$3/2$ representation of the SU(2) group. We choose the commonly-used representation where
\begin{align}
{\mathcal J }_x= 
\begin{pmatrix}
	0 & \frac{\sqrt{3}}{2} & 0 & 0 \\
	\frac{\sqrt{3}}{2} & 0 & 1 & 0 \\
	0 & 1 & 0 & \frac{\sqrt{3}}{2} \\
	0 & 0 & \frac{\sqrt{3}}{2} & 0 
\end{pmatrix} , \quad
{\mathcal J }_y=
\begin{pmatrix}
	0 & \frac{-i \,  \sqrt{3}}{2}  & 0 & 0 \\
	\frac{i \, \sqrt{3}}{2} & 0 & -i & 0 \\
	0 & i & 0 & \frac{-i \, \sqrt{3}}{2}  \\
	0 & 0 & \frac{i \, \sqrt{3}}{2} & 0 
\end{pmatrix}, \quad
{\mathcal J }_z =
\begin{pmatrix}
	\frac{3}{2} & 0 & 0 & 0 \\
	0 & \frac{1}{2} & 0 & 0 \\
	0 & 0 & -\frac{1}{2} & 0 \\
	0 & 0 & 0 & -\frac{3}{2} 
\end{pmatrix}.
\end{align}
The energy eigenvalues, group velocities, BC, and OMM are found to be
\begin{align}
\label{eqeval2}
& \varepsilon_s^{\rm RSW} ( k ) =  s \, v_0 \, k 
\text{ and }
\boldsymbol{v}_s^{\rm RSW} (\mathbf{k}) = 
\nabla_{\mathbf{k}}  \varepsilon_{s}(\mathbf{k})  
=  \frac{ s \, v_0 \,\mathbf k }{k} \,,
\text{ where } 
s \in  \left \lbrace  \pm \frac{1}{2}, \pm \frac{3}{2} \right \rbrace; \nn
& {\mathbf \Omega}^{\chi}_{s}( \mathbf k)  =    
	 - \, \frac{\chi \,   s \,\mathbf k  } {k^3}  \text{ and } 
\mathbf{m}^{\chi}_{s}( \mathbf k) 
= - \frac{e \, \chi\, v_0 \, \mathcal{G}_s  \,\mathbf k } {k^2} \, ,
\text{ where } \lbrace \mathcal{G}_{\pm 1/2}, \,\mathcal{G}_{\pm 3/2} \rbrace
 = \left \lbrace \frac{7}{4}, \, \frac{3}{4} \right \rbrace .
\end{align}

For the case depicted in Fig.~\ref{figdis2}, let us consider the internode-induced conductivity for the chiral quasiparticles at the RSW node with $\chi =1 $. Again, let us assume that the $\chi=-1$ node has an energy offset of $ -\,\Delta $ (where $\Delta > 0$) with respect to the $\chi= 1$ node. Then, application of Eq.~\eqref{eqiso} leads to the node-specific conductivity-tensor components of
\begin{align}
\left( {\sigma}^{1, \rm inter }_{s = 1/2} \right)_{ij}  
=  \left( {\sigma}^{1, \rm inter }_{s =3/2} \right)_{ij}  
 & = 
 \frac { 75  \, e^4  \, v_0^3 \, B_i \, B_j}
{  224 \, \pi^2 \, E_F^2 }
\left[
  \frac {54\, v_0^3
       \left ( E_F + \Delta \right)^2\tau_G}
    {E_F^2\left (112 \, \tilde {v}_0^3 + 27\, v_0^3 \right)
 + 27\, \Delta \, v_0^3 \left( 2\, E_F + \Delta \right)
}  
 - \tau  \right ]
\text{ at the $\Gamma$-point and}\nn
 2\times \left( {\sigma}^{-1, \rm inter }_{s = 1} \right)_{ij}  
 & =   
\frac { 49  \, e^4  \, \tilde{v}_0^3 \, B_i \, B_j}
{ 36 \, \pi^2 \left( E_F + \Delta \right )^2 } 
\left[ \frac {224\, E_F^2 \, \tilde {v}_0^3\, \tau  _G}
 {E_F^2\left (112 \, \tilde {v}_0^3 + 27\, v_0^3 \right)
 + 27\, \Delta \, v_0^3 \left( 2\, E_F + \Delta \right)
 } 
       - \tau  \right ]
\text{ at the $R$-point}.  
\end{align} 
We find that the values for both the RSW bands (with $s=1/2$ and $s= 3/2$) are the same, as expected. Here, $v_0$ and $\tilde v_0 $ denote the group velocities of the pseudospin-3/2 (for the RSW node) and pseudospin-1 (for TSM) quasiparticles, respectively. Summing over the two nodes, the total value is obtained from
$
 \left( {\sigma}^{1, \rm inter }_{1/2} \right)_{ij}  
+  \left( {\sigma}^{1, \rm inter }_{3/2} \right)_{ij}  +
 2\times \left( {\sigma}^{-1, \rm inter }_{1} \right)_{ij}  \,.
$

\section{Summary and future perspectives}
\label{secsum}

In this paper, we have derived a generic expression for the components of the chiral conductivity when we have multifold band-degeneracies. Since the sum of all the monopole charges in the BZ is constrained to vanish, the nodes appear in pairs of $\chi =\pm 1$. Consequently, the presence of band-crossing degeneracies of order higher than two provide a richer playground --- the pair of conjugate nodes in question comprises bands which (1) can be of the same pseudospin variety, carrying the same BC profiles (modulo an overall minus sign); or (2) can carry quantum numbers of two distinct pseudospin-representations (thus, automatically leading to distinct BC profiles). Covering both these two possibilities, we have applied our derived formula to chiral crystals which harbour nodes of the TSM and RSW varieties, thus showing the explicit final expressions of the components of the chiral-conductivity tensor. In our computations, we have accounted for the effects of both the BC and the OMM, thus covering all the topologically-induced modifications in the derivation leading to the linearized Boltzmann equations.
We would like to emphasize that since we have used the methodology based on the relaxation-time approximation, in the future, it will be worthwhile to derive the chiral linear response by going beyond the phenomenological approximations of the relaxation processes \cite{timm}. Due to various contemporary experiments exploring the conductivity of multifold fermions \cite{claudia-multifold}, we expect these theoretical studies to contribute towards a deeper understanding of semimetals with nontrivial topology in their bandstructures.

\section*{Acknowledgments}
We thank Firdous Haidar and Rachika Soni for useful discussions.

\appendix

\section{Linear response from semiclassical Boltzmann equations}
 \label{appboltz}

In this appendix, we review the semiclassical Boltzmann formalism \cite{mermin, ips-kush-review, ips_rahul_ph_strain, ips-rsw-ph}, which is the used to determine the transport coefficients in the regime of linear response. There exists an externally-applied magnetic field $\mathbf B$, which we assume to be small in magnitude, leading to a small cyclotron frequency $\omega_c=e\,B/ m^* $ (where $m^* $ is the effective mass with the magnitude $\sim 0.11 \, m_e$ \cite{params2}, with $m_e$ denoting the electron mass). This allows us to ignore quantized Landau levels, with the regime of validity of our approximations given by $  \omega_c \ll \mu$, where $\mu$ is the Fermi level [i.e., the energy at which the chemical potential cuts the energy band(s)]. Furthermore, we will derive the expressions following from a relaxation-time approximation for the collision integral, which involves using a momentum-independent relaxation time. This implies that we will treat it as a phenomenological parameter. For the intranode scatterings in the collision integrals, we consider the corresponding relaxation time $\tau$. Below, we focus on the transport for a node with chirality $\chi$. The derivation here closely follows the arguments outlined in Refs.~\cite{ips-kush-review, ips_rahul_ph_strain, ips-ruiz, ips-rsw-ph}.

For a 3d system, we define the distribution function for the fermionic quasiparticles occupying a Bloch band labelled by the index $s$ at the node $\chi$, with the crystal momentum $\mathbf k$ and dispersion $\varepsilon_s (\mathbf k)$, by $ f_s^\chi ( \mathbf r , \mathbf k, t) $. Then
\begin{align}
dN_s^\chi = g_s \,f_s^\chi ( \mathbf r , \mathbf k, t) \,
\frac{ d^3 \mathbf k}{(2\, \pi)^3 } \,d^3 \mathbf r
\end{align}
is the number of particles occupying an infinitesimal phase-space volume of $ dV_p = \frac{ d^3 \mathbf k}{(2\, \pi)^3 } 
\,d^3 \mathbf r $, centered at $\left \lbrace \mathbf r , \mathbf k \right \rbrace $ at time $t$. Here, $g_s$ denotes the degeneracy of the band.
In the presence of a nontrivial topology in the bandstructure, a nonzero orbital magnetic moment (OMM) is induced, and there appears a Zeeman-like correction to the energy due to the OMM, which we denote by $\eta_s^\chi (\mathbf k)$. Hence, we define the OMM-corrected dispersion and the corresponding modified Bloch velocity as
\begin{align} 
\label{eqtoten1}
	\xi^{\chi}_s ({\mathbf k})  = \varepsilon_{s}(\mathbf{k})  +  \eta^{\chi}_{s}(\mathbf{k}) 
\text{ and } \boldsymbol{w}^{\chi}_{s}(\mathbf{k})= 
 \nabla_{\mathbf{k}}   \varepsilon_{s}(\mathbf{k})  
 + \nabla_{\mathbf{k}}\eta^{\chi}_{s}(\mathbf{k}) ,
\end{align}
respectively. The Hamilton's equations of motion for the quasiparticles, under the influence of static electric ($\mathbf{E}$) and magnetic ($\mathbf{B}$) fields, are given by \cite{mermin, sundaram99_wavepacket, li2023_planar}
\begin{align}
\label{eqrkdot}
\dot {\mathbf r} &= \nabla_{\mathbf k} \, \xi^{\chi}_{s}  
- \dot{\mathbf k} \, \cross \, \mathbf \Omega^{\chi}_{s} 
 \text{ and }  
\dot{\mathbf k} = -\, e  \left( {\mathbf E}  + \dot{\mathbf r} \, \cross\, {\mathbf B} 
	\right ) \nn
\Rightarrow & \, \dot{\mathbf r}  = \mathcal{D}^{\chi}_{s} 
	\left[   \boldsymbol{w}^{\chi}_{s} + e \left ({\mathbf E}  \cross  
	 \mathbf \Omega^{\chi}_{s} \right )  + e   \left ( \mathbf \Omega^{\chi}_{s} \cdot 
	  \boldsymbol{w}^{\chi}_{s} \right  )  \mathbf B  \right] \text{ and }
\dot{\mathbf k}  = -\, e \,\mathcal{D}^{\chi}_{s} \,  \left[   {\mathbf E} 
+  \left (      \boldsymbol{w}^{\chi}_{s}   \cross  {\mathbf B} \right ) 
+  e  \left (  {\mathbf E}\cdot  {\mathbf B} \right )  \mathbf \Omega^{\chi}_{s}  \right].
\end{align}
where $-\, e$ is the charge carried by each quasiparticle.
Furthermore,
\begin{align}
\mathcal{D}^{\chi}_{s} =  \frac{1}
{ 1 + \, e  \left( \mathbf{B} \cdot\mathbf \Omega^{\chi}_{s}\right) }
\end{align} 
is the factor which modifies the phase-space volume element from $dV_p $ to $  (\mathcal{D}^{\chi}_{s})^{-1} \, dV_p$, such that the Liouville’s theorem (in the absence of collisions) continues to hold in the presence of a nonzero BC \cite{son13_chiral, xiao05_berry, duval06_Berry, son12_berry}.

\subsection{Solution in the absence of internode scattering} 
\label{appintra}

The Fermi-Dirac distribution function,
\begin{align}
\label{eqf0}
f^{(0)}_{s, \chi} (\mathbf r,\mathbf k) \equiv 
	f_{(0)} \big (\xi^\chi_s(\mathbf k) , \mu_\chi, T (\mathbf r) \big )
= \frac{1}
{ 1 + \exp [ \frac{ \xi^\chi_s(\mathbf k)-\mu_\chi} 
{ T (\mathbf r) }  ]}\,,
\end{align} 
describes a local equilibrium situation at the subsystem centred at position $\mathbf r$, at the local temperature $T(\mathbf r )$, and with a spatially uniform chemical potential $\mu_\chi$.
We consider the situation where $T$ and $\mu_\chi $ are constants.
In order to obtain a solution to the full Boltzmann equation for small $ |\mathbf E| $, we assume a small deviation, $\delta  f_s^\chi(\mathbf k)$, from the equilibrium distribution. We have not included any explicit time or spatial dependence in it since $\mathbf E $ is static. Hence, the nonequilibrium time-independent distribution function can be expressed as
\begin{align}
f_s^\chi(\mathbf r,\mathbf k, t) \equiv  f_s^\chi(\mathbf k)
	=  f_{(0)} (\xi^\chi_s(\mathbf k)) +  \delta  f_s^\chi(\mathbf k)\,,
\end{align} 
where we have suppressed showing explicitly the dependence of $f_{(0)} $ on $\mu_\chi $, and $T $. At this point, the magnetic field is not assumed to be small, except for the fact that it should not be so large that the energy levels of the systems get modified by the formation of discrete Landau levels.

To unclutter notations, we will use the superscript ``prime'' to denote differentiation with respect to the variable shown within the brackets [for example, $ f_{(0)}^\prime (u) \equiv \partial_u f_{(0)} (u)$].
We work in the linearized approximation (i.e., we keep terms upto the linear order in the ``smallness parameter''), which implies
\begin{align}
\nabla_{\mathbf k}  f_{(0)}  (\xi^\chi_s(\mathbf k))
=   {\boldsymbol w}_s^\chi 
\,   f_{(0)}^\prime  (\xi^\chi_s(\mathbf k)) \,,
\end{align}
assuming that $\delta f_s^\chi$ is of the same order of smallness as the external perturbation $\mathbf E $. This leads to
the \textit{linearized Boltzmann equation}, given by
\begin{align}
\label{eqkin5}
& - e\, {\mathcal D}_s^\chi  \left [
\left \lbrace {\boldsymbol{w}}_s^\chi 
+ e \left(
{\mathbf \Omega}^\chi_{s} \cdot {\boldsymbol{w}}_s^\chi   \right)  \mathbf B \right \rbrace
\cdot  \mathbf E \right] 
f_{(0)}^\prime  (\xi^\chi_s(\mathbf k)) 
+  e \,  {\mathcal D}_s^\chi \, {\mathbf B} \cdot
\left( {\boldsymbol{w}}_s^\chi   \cross \nabla_{\mathbf k} \right) 
 \delta f_s^\chi (\mathbf k)
 = I_{\rm coll}\,,\quad I_{\rm coll} = - \frac{\delta f_s^\chi (\mathbf k)}{\tau}\,.
\end{align}
Here, $ I_{\rm coll} $ represents the collision integral, which we parametrize by using the phenomenological relaxation time $\tau$. We want to solve the above equation for our planar Hall configurations by using an appropriate ansatz for $\delta f_s^\chi (\mathbf k)$. 

We define the Lorentz-force operator as 
\begin{align}
\check{L} = (\boldsymbol{w}^{\chi}_{s} \cross \mathbf{B}) \cdot \nabla_{\mathbf{k}}\,,
\end{align} 
such that Eq.~\eqref{eqkin5} can be rewritten as
\begin{align}
\label{eqkin6}
&  e\,\mathcal{D}^{\chi}_{s} \left [
\left \lbrace {\boldsymbol{w}}_s^\chi 
+ e \left(
{\mathbf \Omega}^\chi_{s} \cdot {\boldsymbol{w}}_s^\chi   \right)  \mathbf B \right \rbrace
 \cdot \mathbf{E} 
\right ] 
\left[ - f_0^\prime  (\xi^\chi_s(\mathbf k))  \right ]
- e \,\mathcal{D}^{\chi}_{s} \, \check{L} \,{\tilde g}^{\chi}_s (\mathbf{k})
 \delta f_s^\chi (\mathbf k)
=  - \frac{\delta f_s^\chi (\mathbf k)}{\tau} \,.
\end{align}
The solutions and the resulting conductivity tensors for various semimetals have been extensively studied in our earlier papers \cite{ips_rahul_ph_strain, ips-serena, rahul-jpcm, ips-ruiz, ips-rsw-ph, ips-shreya, ips-tilted}.

\subsection{Solution in the presence of internode scattering} 
\label{appinter}

We now discuss how to include internode scatterings in a relaxation-time approximation, where we treat the internode-scattering time $\tau_G$ as a phenomenological constant (analogous to $\tau$). To start with, let us assume that initially, in the infinite past (denoted by time $ t = -\infty $), a pair of conjugate nodes had the same chemical potential $E_F$, characterized by the distribution function
$
f_0 (\varepsilon  ) =\frac{1} {1 + e^{\frac{\varepsilon - E_F} {T}}} \,,
$ 
in the absence of any externally applied fields.
Eventually, on applying the electromagnetic fields, there is the onset of the chiral anomaly, causing each of the two nodes to acquire a local equilibrium value of chemical potential, given by $\mu_\chi$. Therefore, the local equilibrium distribution function at each node is given by
\begin{align}
f^\chi_{s, L} \simeq  f_0 (\xi^{\chi}_{s}) 
 +  \left [- f_{0}^\prime (\xi^{\chi}_{s}) \right ]
 \delta \mu_\chi \,,  \quad
 \delta \mu_\chi \equiv  \mu_\chi - E_F \,.
\end{align}
The intranode-part of the collision integrals ($I_{\text{coll}}^{\text{intra}}$) tend to drive the quasiparticle distribution towards this value, resulting in
\begin{align}
I_{\text{coll}}^{\text{intra}} = - \, 
\frac{f^{\chi}_{s} (\mathbf{k}) 
-  f^\chi_{s, L} }   {\tau } \,.
\end{align}
On the other hand, the internode-part of the collision integrals ($I_{\text{coll}}^{\text{inter}}$) tends to relax the quasiparticle distribution towards the global equilibrium of the chemical potential, $\mu_G $, given by
\begin{align}
& f^\chi_{s, G} \simeq  f_0 (\xi^{\chi}_{s}) 
 +  \left [- f_{0}^\prime (\xi^{\chi}_{s}) \right ]
 \delta \mu_G  \,, \quad
\mu_G =\frac{\mu_\chi + \mu_{-\chi} } {2} \,,\quad
 \delta \mu_G \equiv  \mu_G - E_F = 
 \frac{\delta \mu_\chi + \delta \mu_{-\chi} } {2} \,.
\end{align}
Hence, we get
\begin{align}
I_{\text{coll}}^{\text{inter}} = - \, 
\frac{f^{\chi}_{s} (\mathbf{k}) -  f^\chi_{s, G} }   {\tau_G } \,.
\end{align}

To derive the coefficients of linear response, we parametrize the nonequilibrium 
distribution function as \cite{deng2019_quantum}
\begin{align} 
\label{eqfpar}
f^{\chi}_{s} (\mathbf{k})   & =  
f_{0} (\xi^{\chi}_{s}) + \left [- f_{0}^\prime (\xi^{\chi}_{s}) \right ] 
{\tilde g}^{\chi}_s  (\mathbf{k})\,,
\end{align} 
where $ {\tilde g}^{\chi}_s$ quantifies the deviation of the chemical potential caused by the external probe fields (which are assumed to be spatially uniform and time-independent). With this definition, the total of the collision terms takes the following form:
\begin{align}
I_{\text{coll}} & =  I_{\text{coll}}^{\text{intra}} + I_{\text{coll}}^{\text{inter}}\,, \quad
\frac{ I_{\text{coll}}^{\text{intra}} } { f_{0}^\prime (\xi^{\chi}_{s}) } = 
\frac{ {\tilde g}^{\chi}_s  (\mathbf{k}) 
-  \delta \mu_\chi  }   {\tau } \,,\quad
\frac{ I_{\text{coll}}^{\text{inter}}  }
{ f_{0}^\prime (\xi^{\chi}_{s}) }= 
\frac{ {\tilde g}^{\chi}_s  (\mathbf{k}) 
-  \delta \mu_G  }  {\tau_G }  \,,
\end{align}
Accordingly, $\tau $ and $\tau_G $ represent the phenomenological relaxation-times applicable for the intranode and internode scatterings, respectively. Here, we have assumed the same relaxation time for all the bands involved.

Let us also define the average over all the possible electron states of a physical observable
$\mathcal O^\chi_s ( \xi^\chi_s (\mathbf k) , \mu, T) $ as
\begin{align}
\bar{\mathcal{O}}_\chi \equiv
\left \langle \mathcal{O}^\chi_s ( \xi^\chi_s (\mathbf k), E_F ,T) 
\right  \rangle = 
\frac{
\sum \limits_s \int \frac{d^3 \mathbf{k}} {(2 \pi)^3} 
\left(  {\mathcal D}^\chi_{s}  (\mathbf k) \right)^{-1}
\left [ - f_0^\prime (\xi^\chi_s (\mathbf k) ) \right ] 
\mathcal{O}^\chi_s ( \xi^\chi_s (\mathbf k), E_F ,T) }
{ \sum \limits_{\tilde s} \int 
\frac{d^3 \mathbf{ q} }  {(2 \pi)^3} 
\left( {\mathcal D}^\chi_{\tilde s} (\mathbf q) \right)^{-1}
\left [ - f_0^\prime (\xi^\chi_{\tilde s} (\mathbf{ q}) ) \right ]  } \,.
\end{align}
Since the momentum-integrals run over all the quasiparticle-states at the Fermi level of band $s$ for node $\chi $, the quantity
\begin{align}
\label{eqdoss}
\rho_\chi \equiv   \sum \limits_{s} \int 
\frac{d^3 \mathbf{k} }  {(2 \pi)^3} \left( {\mathcal D}^\chi_{s} (\mathbf k) \right)^{-1}
\left [ - f_0^\prime (\xi^\chi_{s} (\mathbf{k}) ) \right ]
\end{align}
represents the density-of-states at node $\chi $ [cf. Eq.~\eqref{eqdn}].
It naturally follows that the local
equilibrium-distribution function for the chiral quasiparticles is captured by
\begin{align}
 \bar{\tilde{g}}_{\chi}  = \delta \mu_\chi  \,.
\end{align}

For the emergence of the longitudinal magnetoconductivity (LMC), it is necessary that the contribution from intraband scatterings is stronger than that from the interband scattering, implying that we must have $1/\tau  \gg 1/ \tau_G $. Under these conditions, the system first reaches local equilibrium through intraband scattering and, thereafter, achieves global equilibrium through interband scattering \cite{deng2019_quantum}. Thus, the $ \tau  \ll \tau_G $ regime allows us to safely approximate $ {\tilde g}^{\chi}_s \simeq \bar{\tilde{g}}_\chi $ in $I_{\text{coll}}^{\text{inter}}$. Furthermore, the global conservation of the net electric charge gives us the constraint
\begin{align}
\rho_\chi \,\delta \mu_\chi = -\, \rho_{-\chi} \,\delta \mu_{-\chi} \,.
\end{align}
Hence, we finally obtain
\begin{align}
\frac{ I_{\text{coll}}} { f_{0}^\prime (\xi^{\chi}_{s}) } \simeq
\frac{ {\tilde g}^{\chi}_s  (\mathbf{k}) 
-  \delta \mu_\chi  }   {\tau } 
+
\frac{ \delta \mu_\chi - \delta \mu_{-\chi}  }  
{ 2 \, \tau_G } 
= \frac{ {\tilde g}^{\chi}_s (\mathbf{k}) }   
{\tau } 
-
\frac{ \delta \mu_\chi }  
 {  \tau  }  
 \left [  1-\frac{\tau } 
 {2\,\tau_G } \left( 1 + \frac{ \rho_\chi}  {\rho_{-\chi}}
 \right)  \right ] .
\end{align}


On including internode scatterings, the linearized Boltzmann equation, described in Eq.~\eqref{eqkin6}, gets modified to
\begin{align}
\label{eqkininter}
& e\,\mathcal{D}^{\chi}_{s} \left [
\left \lbrace {\boldsymbol{w}}_s^\chi 
+ e \left(
{\mathbf \Omega}^\chi_{s} \cdot {\boldsymbol{w}}_s^\chi   \right)  
\mathbf B \right \rbrace \cdot \mathbf{E} 
\right ]
- e \,\mathcal{D}^{\chi}_{s} \, \check{L} \,{\tilde g}^{\chi}_s (\mathbf{k})
=  -\frac{ {\tilde g}^{\chi}_s (\mathbf{k}) }   
{\tau } 
+ \frac{ \delta \mu_\chi }  
 {  \tau  }  
 \left [  1-\frac{\tau } 
 {2\,\tau_G } \left( 1 + \frac{ \rho_\chi}  {\rho_{-\chi}}
 \right)  \right ]\nn
 \Rightarrow &  
 \left (1 - e \, \tau \, \mathcal{D}^{\chi}_{s} \, \check{L} \right )  
 {\tilde g}^{\chi}_s   (\mathbf{k})
 = -  \,  e\,\tau \, \mathcal{D}^{\chi}_{s}  
\left[ \left \lbrace {\boldsymbol{w}}_s^\chi 
+ e \left({\mathbf \Omega}^\chi_{s} \cdot {\boldsymbol{w}}_s^\chi   \right)  
\mathbf B \right \rbrace \cdot \mathbf{E} \right ] 
 +  \delta \mu_\chi  
 \left [  1-\frac{\tau } 
 {2\,\tau_G } \left( 1 + \frac{ \rho_\chi}  {\rho_{-\chi}}
 \right)  \right ] .
\end{align}
Using the fact that the application of $\check{L} $ on the $ \delta \mu_\chi $-dependent term yields zero,
Eq.~\eqref{eqkininter} is rewritten as
\begin{align}
\label{eqlfinter}
 {\tilde g}^{\chi}_{s} (\mathbf{k}) 
= -\,e\,\tau \sum_{n = 0}^{\infty}
\left (e \, \tau \, \mathcal{D}^{\chi}_{s} \right )^n \check{L}^n 
\left [    \mathcal{D}^{\chi}_{s} \left \lbrace {\boldsymbol{w}}_s^\chi 
+ e \left({\mathbf \Omega}^\chi_{s} \cdot {\boldsymbol{w}}_s^\chi   \right)  
\mathbf B \right \rbrace \cdot \mathbf{E}   \right ]
+    \delta \mu_\chi
 \left [  1-\frac{\tau } 
 {2\,\tau_G } \left( 1 + \frac{ \rho_\chi}  {\rho_{-\chi}}
 \right)  \right ] .
\end{align}
which we solve for $ {\tilde g}^{\chi}_{s} (\mathbf{k})$ recursively.
We can now expand $ {\tilde g}^{\chi}_{s} (\mathbf{k})$ upto any desired order in $ B$, in the limit of weak magnetic field, and obtain the current densities. We observe that $ {\tilde g}^{\chi}_{s} (\mathbf{k})$ consists of two parts, which are of different origins. The first part, which includes the classical effect due to the Lorentz force (given by $n=1$), is independent of $\delta \mu_\chi $. The second term, on the other hand, is proportional to $ \delta \mu_\chi  $ and goes to zero if the values of the chemical potential at the two nodes are the same.

The solution for ${\tilde g}^{\chi}_{s} (\mathbf{k})$, excluding the Lorentz-force part, is obtained from
Eq.~\eqref{eqlfinter} by taking only the $n=0 $ term of the summation on the right-hand side. In other words, we need to consider the equation
\begin{align}
\label{eqinter}
{\tilde g}^{\chi}_{s} (\mathbf{k}) 
= -  \, e \, \tau \, \mathcal{D}^{\chi}_{s} 
\left \lbrace {\boldsymbol{w}}_s^\chi 
+ e \left({\mathbf \Omega}^\chi_{s} \cdot {\boldsymbol{w}}_s^\chi   \right)  
\mathbf B \right \rbrace \cdot \mathbf{E} 
+ \delta \mu_\chi \left [  1-\frac{\tau } 
 {2\,\tau_G } \left( 1 + \frac{ \rho_\chi}  {\rho_{-\chi}}
 \right)  \right ].
\end{align}
First, we need to determine $  \delta \mu_\chi $ self-consistently by taking an average of both the sides of the above equation, which yields
\begin{align}
& \delta \mu_\chi \left [  1-\frac{\tau } 
 {2\,\tau_G } \left( 1 + \frac{ \rho_\chi}  {\rho_{-\chi}}
 \right)  \right ]
=
\bar{\tilde{g}}_{\chi}
+ e \, \tau \left \langle
\mathcal{D}^{\chi}_{s} 
 \left \lbrace {\boldsymbol{w}}_s^\chi 
+ e \left({\mathbf \Omega}^\chi_{s} \cdot {\boldsymbol{w}}_s^\chi   \right)  
\mathbf B \right \rbrace
 \right  \rangle \cdot \mathbf{E}
\Rightarrow  \delta \mu_\chi = 
-\,   \frac{  2 \,  e \, \tau_G}  
{  1 + \frac{ \rho_\chi}  { \rho_{-\chi}} }
\frac{ \boldsymbol {\mathcal I}^\chi \cdot \mathbf{E} }
{\rho_\chi} \,, \text{ where}
\nn &
\boldsymbol {\mathcal I}^\chi  = \rho_\chi
\left \langle
\mathcal{D}^{\chi}_{s} 
\left \lbrace {\boldsymbol{w}}_s^\chi 
+ e \left({\mathbf \Omega}^\chi_{s} \cdot {\boldsymbol{w}}_s^\chi   \right)  
\mathbf B \right \rbrace\right  \rangle 
=
\sum \limits_s \int \frac{d^3 \mathbf{k}} {(2 \pi)^3} 
\left [ - f_0^\prime (\xi^\chi_s (\mathbf k) ) \right ] 
 \left \lbrace {\boldsymbol{w}}_s^\chi 
+ e \left({\mathbf \Omega}^\chi_{s} \cdot {\boldsymbol{w}}_s^\chi   \right)  
\mathbf B \right \rbrace .
\end{align}

The non-anomalous-Hall contribution to the current, excluding the Lorentz-force part, is obtained by using Eqs.~\eqref{eqfpar} and \eqref{eqinter}, leading to
\begin{align}
& {\mathbf {\bar J}}^{\chi }_s = -\, e   \int
\frac{ d^3 \mathbf k} {(2\, \pi)^3 } 
\left [- f_{0}^\prime (\xi^{\chi}_{s}) \right ] 
\left[   \boldsymbol{w}^{\chi}_{s} 
+ e   \, ( \mathbf \Omega^{\chi}_{s} \cdot 
\boldsymbol{w}^{\chi}_{s} ) \, \mathbf B  \right]
{\tilde g}^{\chi}_s  (\mathbf{k})
=
 {\mathbf { J}}^{\chi, \rm intra }_s + {\mathbf { J}}^{\chi, \rm inter }_s \,,
\nn & {\mathbf { J}}^{\chi, \rm intra }_s =
 e^2 \, \tau   \int
\frac{ d^3 \mathbf k}{(2\, \pi)^3 } \,\mathcal{D}^{\chi}_{s}
\left [- f_{0}^\prime (\xi^{\chi}_{s}) \right ] 
\left[   \boldsymbol{w}^{\chi}_{s} 
+ e   \, ( \mathbf \Omega^{\chi}_{s} \cdot 
\boldsymbol{w}^{\chi}_{s} ) \, \mathbf B  \right] 
\left \lbrace {\boldsymbol{w}}_s^\chi 
+ e \left({\mathbf \Omega}^\chi_{s} \cdot {\boldsymbol{w}}_s^\chi   \right)  
\mathbf B \right \rbrace \cdot \mathbf{E} \,,
\nn & {\mathbf { J}}^{\chi, \rm inter }_s =
\frac{  2 \,  e^2 }  
{  1 + \frac{ \rho_\chi}  { \rho_{-\chi}} }  \left [  \tau_G
-\frac{\tau} {2} 
 \left( 1 + \frac{ \rho_\chi}  {\rho_{-\chi}}
 \right)  \right ]
 \int
\frac{ d^3 \mathbf k}{(2\, \pi)^3 } 
\left [- f_{0}^\prime (\xi^{\chi}_{s}) \right ] 
\left[   \boldsymbol{w}^{\chi}_{s} 
+ e   \, ( \mathbf \Omega^{\chi}_{s} \cdot 
\boldsymbol{w}^{\chi}_{s} ) \, \mathbf B  \right]
\frac{\boldsymbol {\mathcal I}^\chi \cdot \mathbf{E} }
{\rho_\chi}  \,.
\end{align}
Therefore, the conductivity tensor, corresponding to the internode-scattering-induced current, is given by
\begin{align}
\label{eqcond11}
\left( {\sigma}^{\chi, \rm inter }_{s} \right)_{ij} =
\frac{  e^2  \, \rho_{-\chi} }  
{ \rho_\chi \, \rho_G  }  
\left [  \tau_G
-\frac{\tau \, \rho_G } { \rho_{-\chi}}   \right ]
 \int
\frac{ d^3 \mathbf k}{(2\, \pi)^3 } 
\left [- f_{0}^\prime (\xi^{\chi}_{s}) \right ] 
\;  \left[   \left ( {w}^{\chi}_{s} \right)_i
+ e   \, ( \mathbf \Omega^{\chi}_{s} \cdot 
\boldsymbol{w}^{\chi}_{s} ) \,  B_i  \right]\,
\, {\mathcal I}_j^\chi  \,, \quad
\rho_G = \frac{ \rho_\chi + \rho_{-\chi} }  {2} \,.
\end{align}

\section{Terms expanded upto order $B^2$}
\label{appexp}

In this appendix, we expand the integrand of Eq.~\eqref{eqcond1} by keeping terms upto order $ B^2 $.
For this purpose, we define
\begin{align}
& \rho_\chi  =
\rho^{(0)}_\chi + \rho^{(1)}_\chi + \rho^{(2)}_\chi + \order{B^3}\,,\nn
&  \rho^{(0)}_\chi = \sum_s \int
\frac{ d^3 \mathbf q} {(2\, \pi)^3 }   
\left \lbrace  -  f_0^\prime  
\big (  \varepsilon_s^\chi  (\mathbf  q) \big) \right \rbrace\,,\quad
\rho^{(1)}_\chi = \sum_s \int
\frac{ d^3 \mathbf q} {(2\, \pi)^3 }  
\left[
e \,\mathbf {B}\cdot\mathbf{ \Omega}_s^{\chi} (\mathbf  q) 
\,  f_0^\prime  
\big (  \varepsilon_s^\chi  (\mathbf  q) \big) 
+
\eta_s^{\chi} (\mathbf  q)  
\left \lbrace  -  f_0^{\prime \prime}  
\big (  \varepsilon_s^\chi  (\mathbf  q) \big) \right \rbrace
\right ],
\nn & \rho^{(2)}_\chi = \sum_s \int
\frac{ d^3 \mathbf q} {(2\, \pi)^3 } 
 \left[
 e^2 \left \lbrace   {\mathbf  B} 
  \cdot {\mathbf \Omega}_s^{\chi} (\mathbf  q)
\right \rbrace^2
\left \lbrace  -  f_0^{\prime}  
\big (  \varepsilon_s^\chi  (\mathbf  q) \big) \right \rbrace
+
  e \left \lbrace   {\mathbf  B} 
  \cdot {\mathbf \Omega}_s^{\chi} (\mathbf  q)
\right \rbrace
 \eta_s^{\chi} (\mathbf  q) 
\, f_0^{\prime \prime}  
\big (  \varepsilon_s^\chi  (\mathbf  q) \big)
+
 \frac{ \left \lbrace \eta_s^{\chi} (\mathbf  q) 
 \right \rbrace ^2 }  {2}
\left \lbrace  -  f_0^{\prime \prime \prime}  
\big (  \varepsilon_s^\chi  (\mathbf  q) \big) \right \rbrace
\right ],
\end{align}
\begin{align}
& \rho_G =   \rho^{(0)}_G + \rho^{(1)}_G + \rho^{(2)}_G +\order{B^3}\,,
\quad \rho^{(0)}_G = \frac{ \rho^{(0)}_\chi + \rho^{(0)}_{-\chi} } {2} \,, \quad
\rho^{(1)}_G   =
\frac{ \rho^{(1)}_\chi +  \rho^{(1)}_{-\chi} } {2} \,, \quad
\rho^{(2)}_G =
\frac{ \rho^{(2)}_\chi +  \rho^{(2)}_{-\chi} } {2} \,,
\end{align}
\begin{align}
& \mathcal I_j^\chi  =
\mathcal I^{\chi, 0}_j + \mathcal I^{\chi, 1}_j  + \mathcal I^{\chi, 2}_j + \order{B^3}\,,\nn
& \mathcal I^{\chi, 0}_j  =  \sum_s \int
\frac{ d^3 \mathbf q} {(2\, \pi)^3 }  \, 
\big ( v_s^\chi (\mathbf  q) \big)_j
\left \lbrace  -  f_0^\prime  \big (  \varepsilon_s^\chi  (\mathbf  q) \big) \right \rbrace\,,\nn &
\mathcal I^{\chi, 1}_j =  \sum_s \int
\frac{ d^3 \mathbf q} {(2\, \pi)^3 }  \left[
\left \lbrace 
  e  \left (  {\mathbf \Omega}_s^{\chi} (\mathbf  q) 
  \cdot  {\boldsymbol  v}_s^\chi (\mathbf q) \right )  B_j 
 + \big ( u_s^\chi (\mathbf  q) \big)_j
\right  \rbrace
 \left \lbrace  -  f_0^{\prime}  
 \big (  \varepsilon_s^\chi  (\mathbf  q) \big) \right \rbrace
+ 
\eta_s^{\chi} (\mathbf  q)\,\big ( v_s (\mathbf  q)\big)_j
\left \lbrace  -  f_0^{\prime \prime}  \big (  \varepsilon_s^\chi  (\mathbf  q) \big) \right \rbrace
\right ],
\nn & \mathcal I^{\chi, 2}_j = \sum_s \int
\frac{ d^3 \mathbf q} {(2\, \pi)^3 }  \,
  \Bigg [ e 
  \left \lbrace {\mathbf \Omega}_s^{\chi} (\mathbf  q )\cdot 
  {\boldsymbol u}_s^{\chi} (\mathbf q) \right \rbrace
 B_j \left \lbrace  -  f_0^{\prime}  
\big (  \varepsilon_s^\chi  (\mathbf  q) \big) \right \rbrace
+
\eta_s^{\chi} (\mathbf q) \,
\left \lbrace 
e\left( {\mathbf \Omega}_s^{\chi} (\mathbf  q) \cdot 
{\boldsymbol  v}^\chi_s (\mathbf  q)
 \right )  B_j
+ \left( u_s^{\chi} (\mathbf  q) \right)_j \right  \rbrace
  \left \lbrace  -  f_0^{\prime \prime}  
  \big (  \varepsilon_s^\chi  (\mathbf  q) \big) \right \rbrace
\nn & \hspace{ 3.5 cm }
+
\frac { \left  (\eta_s^{\chi} (\mathbf  q)\right )^2\,
  \big ( v_s^\chi (\mathbf   q)\big)_j}   {2} 
\left \lbrace  -  f_0^{\prime \prime \prime} 
 \big (  \varepsilon_s^\chi  (\mathbf  q) \big) \right \rbrace    
\Bigg ] \,.
\end{align}
\begin{align}
\label{eqsigexpr}
\left( {\sigma}^{\chi, \rm inter }_{s} \right)_{ij} = 
\frac {e^2  
\left [ \tau \, \rho_G^{(0)} - \tau_G \, \rho_{-\chi}^{(0)} \right ] }
 { \rho_G^{(0)} \, \rho_{\chi}^{(0)} }
\int
\frac{ d^3 \mathbf k} {(2\, \pi)^3 } 
\left [
 \varsigma_{0, ij}^{\chi, s}
+
 \varsigma_{1, ij}^{ \chi, s} 
+
 \varsigma_{ 2, ij}^{ \chi, s} \right ] + \order{B^3} \,,
\end{align}
where
\begin{align}
&  \varsigma_{0, ij}^{\chi, s} (\mathbf k) = -\, \mathcal {I}_j^{\chi, 0}
 \left( v_s^{\chi} \right)_i  
 \left \lbrace  -  f_0^{\prime}  \big (  \varepsilon_s^\chi \big) \right \rbrace,
\end{align}
\begin{align}
 \varsigma_{1, ij}^{\chi, s} (\mathbf k) & =   
\Bigg[  \Big [
  -e \, {\mathbf \Omega}_s^{\chi} \cdot  {\boldsymbol  v}_s^{\chi}\,  B_i 
   + 
\bigg \lbrace
\frac{\tau_G} {\rho_G^{(0)} } \,
\frac {   
 \rho_G^{(0)}\, \rho_{-\chi}^{(1)}
 - \rho_G^{(1)} \, \rho_{-\chi}^{(0)} }
  {  \tau  \, \rho_G^{(0)} - \tau_G \, \rho_{-\chi}^{(0)}
}
   + \frac { \rho_{\chi}^{(1)} }  {\rho_{\chi}^{(0)}}
\bigg  \rbrace 
\left ( v_s^{\chi}  \right)_i
   - \left ( u_s^{\chi} \right)_i
     \Big ] \,   {\mathcal  I}_j^{\chi, 0}
-  \left ( v_s^{\chi} \right)_i \,  {\mathcal  I}_j^{\chi, 1}
\Bigg ]  
 \left \lbrace  -  f_0^{\prime}  \big (  \varepsilon_s^\chi \big) \right \rbrace
\nn & \quad \;- \eta_s^{\chi} 
\left ( v_s^{\chi} \right)_i \, {\mathcal  I}_j^{\chi,  0}
\,\left \lbrace  -  f_0^{\prime \prime}  \big (  \varepsilon_s^\chi \big) \right \rbrace \,,
\end{align}
\begin{align}
\label{eqsig3}
  \varsigma_{2, ij}^{\chi, s} (\mathbf k) & = 
 \Bigg[
e  \, \rho_G^{(0)}  \,  \rho_{\chi}^{(0)} 
\, {\mathbf  \Omega}_s^{\chi} \cdot {\boldsymbol  v}_s^{\chi}
\left  \lbrace
\rho_G^{(0)}  \left (\tau_G \, \rho_{\chi}^{(0)}  
        \rho_{-\chi}^{(1)} 
+ \tau \,  \rho_G^{(0)}  
        \rho_{\chi}^{(1)}  \right)
- \tau_G \, \rho_{-\chi}^{(0)} \left (\rho_G^{(1)} 
        \rho_{\chi}^{(0)}  
+ \rho_G^{(0)} \,   \rho_{\chi}^{(1)}  \right) 
\right \rbrace B_i 
\nn & \hspace{ 0.6 cm }
+ e   \left( \rho_G^{(0)}   \,\rho_{\chi}^{(0)}   \right)^2
{\mathbf \Omega}_s^{\chi} \cdot {\boldsymbol  u}_s^{\chi}
\left (\tau_G  \,\rho_{-\chi}^{(0)}  
- \tau \,  \rho_G^{(0)}   \right)  B_i
- \tau_G \, \rho_{-\chi}^{(0)}  \,  \rho_G^{(0)} \, \rho_{\chi}^{(0)}  
\left \lbrace  
\, \rho_G^{(1)}\, \rho_\chi^{(0)}
-  \rho_G^{(0)}     \, \rho_{\chi}^{(1)} 
\right \rbrace   \left( u_s^{\chi} \right)_i 
+
\tau_G \left( \rho_G^{(0)}  \, \rho_\chi^{(0)}  \right)^2    
\rho_{-\chi}^{(1)}  \left( u_s^{\chi} \right)_i
\nn & \hspace{ 0.6 cm }
+ \tau  \left( \rho_G^{(0)} \right)^3   \rho_{\chi}^{(0)}   
\, \rho_{\chi}^{(1)}  \left( u_s^{\chi} \right)_i
+ \tau_G \, \rho_{-\chi}^{(0)}  
\left( \rho_G^{(1)} \right)^2  
 \left( \rho_\chi^{(0)}   \right)^2 \left( v_s^{\chi} \right)_i
 + 
 \tau_G \,  \rho_{-\chi}^{(0)} 
\left(  \rho_G^{(0)} \right)^2   
 \left( \rho_{\chi}^{(1)} \right)^2 \left( v_s^{\chi} \right)_i
 + 
 \tau_G \, \rho_{-\chi}^{(0)}  \, \rho_G^{(0)}   
\rho_G^{(1)}  \, \rho_{\chi}^{(0)}   
\rho_{\chi}^{(1)}   \left( v_s^{\chi} \right)_i 
 \nn & \hspace{ 0.6 cm }
- \tau_G \, \rho_G^{(0)}   \,
\rho_G^{(1)}   \left( \rho_{\chi}^{(0)} \right)^2
 \rho_{-\chi}^{(1)} \left( v_s^{\chi} \right)_i 
-
\tau \left(  \rho_G^{(0)} \right)^3  
\left( \rho_{\chi}^{(1)}  \right)^2 \left( v_s^{\chi} \right)_i 
- \tau_G
 \left( \rho_G^{(0)}   \right)^2  
  \rho_{\chi}^{(0)}  \, \rho_{-\chi}^{(1)}  \, \rho_{\chi}^{(1)} 
 \left( v_s^{\chi} \right)_i 
-
\tau_G  \,\rho_G^{(2)} 
\rho_{-\chi}^{(0)}  \, \rho_G^{(0)}   
 \left( \rho_\chi^{(0)}   \right)^2 \left( v_s^{\chi} \right)_i 
\nn & \hspace{ 0.6 cm }
+
\tau_G \, \rho_{-\chi}^{(2)} 
\left( \rho_G^{(0)}   \right)^2    
 \left( \rho_\chi^{(0)}   \right)^2 \left( v_s^{\chi} \right)_i
 + 
 \rho_{\chi}^{(2)}  
\left( \rho_G^{(0)}   \right)^2  
\rho_{\chi}^{(0)}  
\left (\tau \,  \rho_G^{(0)}  
   - \tau_G  \rho_{-\chi}^{(0)}  \right) \left( v_s^{\chi} \right)_i 
 \Bigg]
\; \frac { 
{\mathcal  I}_j^{\chi,  0} \,
 \left \lbrace  -  f_0^{\prime}  
 \big (  \varepsilon_s^\chi \big) \right \rbrace }
{\left \lbrace \rho_G^{(0)}\, \rho_{\chi}^{(0)} \right \rbrace^2
\; \left[ \tau \, \rho_G^{(0)} - \tau_G \, \rho_{-\chi}^{(0)} 
\right ] }
\nn \nn & \quad \; +
\Bigg [
-  e \, 
{\mathbf \Omega}_s^{\chi} \cdot {\boldsymbol v}_s^{\chi} \, B_i
-  \left(u_s^{\chi} \right)_i
+ 
 \left \lbrace
\frac{ \tau_G } {\rho_G^{(0)}}  \, 
\frac {\rho_G^{(0)}   \rho_{-\chi}^{(1)}
     - \rho_{-\chi}^{(0)}  \rho_G^{(1)}  }
{ \tau  \,  \rho_G^{(0)} 
    - \tau_G \, \rho_{-\chi}^{(0)}  }
+\frac {\rho_{\chi}^{(1)} }
{\rho_{\chi}^{(0)} } 
 \right \rbrace  \left (v_s^{\chi} \right)_i 
\Bigg ]
\, {\mathcal  I}_j^{\chi,  1} \, 
 \left \lbrace  -  f_0^{\prime}  
 \big (  \varepsilon_s^\chi \big) \right \rbrace 
 -   \left (v_s^{\chi} \right)_i   \,
{\mathcal  I}_j^{\chi,  2}  
 \left \lbrace  -  f_0^{\prime}  
 \big (  \varepsilon_s^\chi \big) \right \rbrace 
\nn & \quad \;  
+ \eta_s^{\chi} \Bigg[  
\Big [
  -e \, {\mathbf \Omega}_s^{\chi} \cdot  {\boldsymbol  v}_s^{\chi}\,  B_i
- \left ( u_s^{\chi} \right)_i   + 
\bigg \lbrace
\frac{\tau_G} {\rho_G^{(0)} } \,
\frac {   
 \rho_G^{(0)}\, \rho_{-\chi}^{(1)}
        - \rho_G^{(1)} \, \rho_{-\chi}^{(0)} }
  {  \tau  \, \rho_G^{(0)} - \tau_G \, \rho_{-\chi}^{(0)}
}
   + \frac { \rho_{\chi}^{(1)} }  {\rho_{\chi}^{(0)}}
\bigg  \rbrace 
\left ( v_s^{\chi}  \right)_i
     \Big ] \,   {\mathcal  I}_j^{\chi, 0}
-  \left ( v_s^{\chi} \right)_i \,  {\mathcal  I}_j^{\chi, 1}
\Bigg ]
 \left \lbrace  -  f_0^{\prime \prime}  
 \big (  \varepsilon_s^\chi \big) \right \rbrace
\nn & \quad \; -  \frac{ 
\left( \eta_s^{\chi} \right)^2
\left ( v_s^{\chi} \right)_i \, {\mathcal  I}_j^{\chi,  0}} {2}
\left \lbrace  -  f_0^{\prime \prime \prime}  
\big (  \varepsilon_s^\chi \big) \right \rbrace.
\end{align}
In the last three equations, we have suppressed the $\mathbf k$-dependence of the various quantities.

For the cases when $ \varepsilon^\chi_s $ is a function of magnitudes of the momentum components [i.e., $ \varepsilon^\chi_s (\mathbf k) =  \varepsilon^\chi_s (|k_x|, |k_y|, |k_z|)$], $ f_0^{\prime}\big (  \varepsilon_s^\chi (\mathbf k) \big) $ and its derivatives are even functions of $\mathbf k $. Since integrands which are odd functions of the momentum components must vanish, we conclude that $ \rho^{(1)}_\chi  =  \rho^{(1)}_G = I^{\chi, 0}_j = \int
\frac{ d^3 \mathbf k} {(2\, \pi)^3 } \,
 \varsigma_{0, ij}^{\chi, s} = \int \frac{ d^3 \mathbf k} {(2\, \pi)^3 } \,
 \varsigma_{1, ij}^{\chi, s}   = 0 $. Furthermore, 
\begin{align}
\mathcal I^{\chi, 1}_j &=  \sum_s \int \frac{ d^3 \mathbf k} {(2\, \pi)^3 } 
 \left[ 
   e \, {\mathbf \Omega}_s^{\chi} \cdot {\boldsymbol v}_s^{\chi} \,B_j \,
 \left \lbrace  -  f_0^{\prime}  \big (  \varepsilon_s^\chi \big) \right \rbrace 
+  \mathbf{m}^{\chi}_s \cdot {\mathbf B}   \left ( v_s^{\chi} \right)_j
  f_0^{\prime \prime}  \big (  \varepsilon_s^\chi \big) 
 \right ] 
 =
 \sum_{ s} \Upsilon^{\chi,  s}_j ,\nn 
\Upsilon^{\chi, s}_j & =   B_j \int
\frac{ d^3 \mathbf k} {(2\, \pi)^3 } 
\left [
 e \, {\mathbf \Omega}_s^{\chi} (\mathbf k) 
 \cdot  {\boldsymbol  v}_s^{\chi} (\mathbf k) 
\left \lbrace - 
 f_0^\prime \big (\varepsilon_s^{\chi} (\mathbf k) \big)  
 \right  \rbrace
+
  \left ( m_s^{\chi} (\mathbf k) \right)_j  
  \left (  v_s^{\chi} (\mathbf k) \right)_j
f_0^{\prime \prime} \big (\varepsilon_s^{\chi} (\mathbf k)  \big) 
\right ] ,
\end{align}
and Eq.~\eqref{eqsig3} simplifies to
\begin{align}
  \varsigma_{2, ij}^{\chi, s} (\mathbf k) & = 
\left [
-  e \, 
{\mathbf \Omega}_s^{\chi} \cdot {\boldsymbol v}_s^{\chi} \, B_i
-  \left(u_s^{\chi} \right)_i
+ 
 \left \lbrace
\frac{ \tau_G } {\rho_G^{(0)}}  \, 
\frac {\rho_G^{(0)}   \rho_{-\chi}^{(1)}
     - \rho_{-\chi}^{(0)}  \rho_G^{(1)}  }
{ \tau  \,  \rho_G^{(0)} 
    - \tau_G \, \rho_{-\chi}^{(0)}  }
+\frac {\rho_{\chi}^{(1)} }
{\rho_{\chi}^{(0)} } 
 \right \rbrace  \left (v_s^{\chi} \right)_i 
\right ]
 {\mathcal  I}_j^{\chi,  1} \, 
 \left \lbrace  -  f_0^{\prime}  \big (  \varepsilon_s^\chi \big) \right \rbrace 
 -   \left (v_s^{\chi} \right)_i   \,
{\mathcal  I}_j^{\chi,  2}  
 \left \lbrace  -  f_0^{\prime}  \big (  \varepsilon_s^\chi \big) \right \rbrace 
\nn & \qquad \;  
- \eta_s^{\chi}   \left ( v_s^{\chi} \right)_i \,  {\mathcal  I}_j^{\chi, 1}
 \left \lbrace  -  f_0^{\prime \prime}  \big (  \varepsilon_s^\chi \big) 
 \right \rbrace,
\end{align}
with
\begin{align}
 \int \frac{ d^3 \mathbf k} {(2\, \pi)^3 } 
\,  \varsigma_{2, ij}^{\chi, s} (\mathbf k) & = 
  \int \frac{ d^3 \mathbf k} {(2\, \pi)^3 } 
 \left[ 
  \left (-  e \, {\mathbf \Omega}_s^{\chi} \cdot {\boldsymbol v}_s^{\chi} 
\right ) B_i\,
 {\mathcal  I}_j^{\chi,  1} \, 
 \left \lbrace  -  f_0^{\prime}  \big (  \varepsilon_s^\chi \big) \right \rbrace 
- \eta_s^{\chi}   \left ( v_s^{\chi} \right)_i \,  {\mathcal  I}_j^{\chi, 1}
 \left \lbrace  -  f_0^{\prime \prime}  \big (  \varepsilon_s^\chi \big) 
 \right \rbrace \right ]
\nn & =  {\mathcal  I}_j^{\chi, 1}
  \int \frac{ d^3 \mathbf k} {(2\, \pi)^3 } 
 \left[ 
  \left (-  e \, {\mathbf \Omega}_s^{\chi} \cdot {\boldsymbol v}_s^{\chi} \right ) B_i\,
 \left \lbrace  -  f_0^{\prime}  \big (  \varepsilon_s^\chi \big) \right \rbrace 
+ \left( m_s^\chi \right)_i \, B_i
\left ( v_s^{\chi} \right)_i 
 \left \lbrace  -  f_0^{\prime \prime}  \big (  \varepsilon_s^\chi \big) 
 \right \rbrace \right ]
 = -\, \Upsilon^{\chi, s}_i \, {\mathcal  I}_j^{\chi,  1} .
\end{align}
Plugging these in Eq.~\eqref{eqsigexpr}, we get the final form as
\begin{align}
& \left( {\sigma}^{\chi, \rm inter }_{s} \right)_{ij} = 
\frac {e^2  
\left [ \tau_G \, \rho_{-\chi}^{(0)} 
- \tau \, \rho_G^{(0)} \right ]
\Upsilon^{\chi, s}_i \, {\mathcal  I}_j^{\chi, 1} }
 { \rho_G^{(0)} \, \rho_{\chi}^{(0)} } 
 +\order{B^3} \,,
\end{align}
leading to the expression shown in Eq.~\eqref{eqiso} of the main text.

\section{Useful integrals}
\label{secint}

In the main text, we have to deal with integrals of the form
\begin{align}
\mathcal I = \int \frac{d^3 {\bf{k}} }{(2 \pi )^{3}} \, F ({\bf{k}} , \varepsilon_{s} ) \,
f_{0}^\prime ( \varepsilon_{s} )  \,, \text{ where }
\varepsilon_{s} =  s \, v_0\, k \,.
\end{align}
Clearly, it is convenient to switch to the spherical polar coordinates, using the standard transformations comprising
\begin{align}
\label{eqpolar} 
k_x = \frac{ \tilde  \varepsilon \cos \phi \sin \theta} 
 {  s\, v_0}\,, \quad
 k_y = 
\frac{\tilde \varepsilon \sin \phi \sin \theta} { s\, v_0}\,, 
\quad k_{z} 
 = \frac{ \tilde  \varepsilon \cos  \theta} { s\, v_0} \,,
\end{align}
where $\tilde  \varepsilon \in [0, \infty )$, $\phi \in [0, 2 \pi )$, and $\theta \in [0, \pi ]$. The Jacobian of the transformation is $\mathcal{J} (\tilde  \varepsilon , \theta ) =   \frac{ \tilde  \varepsilon^2  \sin \theta}
{ s^3 \, v_0^3} $. This leads to
\begin{align}
\int_{- \infty}^{ \infty} d^3 \mathbf{k } \rightarrow   
\int_{0}^{ \infty} d  \tilde  \varepsilon  \int_{0}^{ 2 \,\pi} d \phi 
\int_{0}^{  \pi} d \theta \,   \mathcal{J} ( \tilde  \varepsilon , \theta ) 
\text{ and } 
 \varepsilon_{s}(\mathbf{k})  \rightarrow    \tilde  \varepsilon \,,
\end{align}
Since the dispersion does not depend on $\theta$ or $\phi$, we can immediately perform the angular integrals as the first step, which makes many terms disappear.


\bibliography{ref_inter}

\end{document}